\documentclass[12pt]{article}
\usepackage{fullpage}
\usepackage{epsfig}
\usepackage{amsmath}
\usepackage{amssymb}
\usepackage{color}

\begin{document}

\vspace{0.5 cm}

\centerline{\bf\Large Lattice study of light scalar tetraquarks with ${\bf  I=0,2,\tfrac{1}{2},\tfrac{3}{2}}$ : }

\vspace{0.3 cm}

\centerline{\bf\Large are $\sigma$ and $\kappa$ tetraquarks?}

\vspace{1 cm}

\centerline{\bf Sasa Prelovsek$^{(a)}$, Terrence Draper$^{(b)}$, Christian B. Lang$^{(c)}$, Markus Limmer$^{(c)}$,}

\vspace{0.3 cm}

 \centerline{\bf  Keh-Fei Liu$^{(b)}$, Nilmani Mathur$^{(d)}$ and Daniel Mohler$^{(e)}$ }

\vspace{1cm}

\centerline{\it (a) Department of Physics, University of Ljubljana and Jozef Stefan Institute, Ljubljana, Slovenia.}

\vspace{0.2cm}

\centerline{\it (b) Department of Physics and Astronomy, University of Kentucky, Lexington, KY 40506, USA.}

\vspace{0.2cm}

\centerline{\it (c) Institut f\"{u}r Physik, FB Theoretische Physik, Universit\"{a}t Graz, A-8010 Graz, Austria.}

\vspace{0.2cm}

\centerline{\it (d) Department of Theoretical Physics, Tata Institute of Fundamental Research, Mumbai, India.}

\vspace{0.2cm}

\centerline{\it (e) TRIUMF, 4004 Wesbrook Mall Vancouver, BC V6T 2A3, Canada}

\vspace{0.3cm}

\centerline{e-mail: {\tt sasa.prelovsek@ijs.si}}

\vspace{1cm}

{\bf Abstract}

\vspace{0.2cm}

We investigate whether the lightest scalar mesons $\sigma$ and  $\kappa$ have a large   tetraquark component $\bar q\bar q qq$, as is strongly supported by many phenomenological studies.  A search  for possible light tetraquark states with $J^{PC}=0^{++}$ and $I=0,~2,~1/2,~3/2$ on the lattice is presented. We perform the two-flavor dynamical simulation with Chirally Improved quarks and the quenched simulation with overlap quarks, finding qualitative agreement between both results.  
The spectrum is determined using the generalized eigenvalue method with a number of tetraquark interpolators at the source and the sink, and we omit the disconnected contractions. The time-dependence of the eigenvalues at finite temporal extent of the lattice is explored also  analytically. 
In all the channels, we unavoidably find lowest  scattering states $\pi(k)\pi(-k)$ or $K(k)\pi(-k)$ with back-to-back momentum $k=0,~2\pi /L~,..~$. However, we find an additional light state in the $I=0$ and $I=1/2$ channels, which may be interpreted as  the observed resonances $\sigma$ and $\kappa$ with a sizable tetraquark component.  
In the exotic repulsive channels $I=2$ and $I=3/2$, where no resonance is observed, we find no light state in addition to the scattering  states.

\newpage

\section{Introduction}

The only well established hadron states so far are mesons $\bar qq$ and baryons $qqq$. No exotic states like tetraquark $[\bar q\bar q] [qq]$,  pentaquark $\bar qqqqq$, hybrid $\bar qqG$  or molecular $(\bar qq)(\bar qq),~(\bar qq)(qqq)$  have  been confirmed beyond doubt, although there are several serious candidates in the light and hidden charm sectors. 
   Perhaps the most prominent tetraquark candidate is the $Z^+(4430)$ resonance, discovered by Belle \cite{Zbelle}: it decays to $\pi^+\psi^\prime$, so it must have a minimal quark content $\bar du\bar c c$, but it has not been confirmed by Babar  \cite{Zbabar}. 

It is still not established whether the lightest scalar mesons  $\sigma$,  $\kappa$, $a_0(980)$ and $f_0(980)$  are conventional  $\bar qq$ states or they have important $\bar q \bar qqq$ or glue Fock components. A sizable glue component of isoscalar $\sigma$ is supported by some phenomenological studies \cite{sigma_glue}, but we will not explore this Fock component in this work. 
We will focus on the  $\bar q \bar qqq$ Fock component, which arises in the case of a tetraquark $[qq][\bar q\bar q]$  or  in the case of  a mesonic molecule $(\bar qq)(\bar qq)$.  The tetraquarks $[qq][\bar q\bar q]$ are composed of a scalar diquark ($\bar 3_{C,F}$) and anti-diquark ($3_{C,F}$) in $L=0$; they form a flavor nonet and are expected to be light \cite{tetra_phenomenology,pelaez}. A mesonic molecule $(\bar qq)(\bar qq)$ is composed of two color-singlet mesons ($\pi,K$) held together by pion exchange \cite{molecules}. 
Both $\bar q \bar qqq$ interpretations expect that the $I=1$ state 
 ($\bar u\bar s sd$)  is heavier than the $I=1/2$ state ($\bar u\bar d ds$) due to $m_s>m_d$, in agreement with experimental ordering $m_{a_0(980)}>m_{\kappa}$.   On the other hand,  the conventional $\bar ud$ and $\bar us$ states can hardly explain the observed mass ordering. Both $\bar q\bar qqq$ interpretations also naturally explain the large observed coupling of $a_0(980)$ and $f_0(980)$ to $\bar KK$, which is due to the additional valence pair $\bar ss$.

  In this paper we use a lattice QCD simulation to address the question whether the lightest scalar mesons $\sigma$ ($I=0$) and $\kappa$ ($I=1/2$) have a sizable $\bar q\bar qqq$ component. The quantities studied in our present  simulation do not distinguish between a tetraquark and a mesonic molecule. When we use a word ``tetraquark'' below, we have in mind  both types of exotic $\bar q\bar qqq$ states.

The $\sigma$ resonance is now widely accepted since its pole with $m_\sigma=441{+16\atop -8}$ MeV and 
$\Gamma_\sigma=544{+18 \atop -25}$ MeV was determined  in a model-independent way \cite{leutwyler}.  The $\kappa$ resonance pole with $m_\kappa=658\pm 13$ MeV and $\Gamma_\kappa=557\pm 24$ MeV   was determined in a similar manner  \cite{descotes}. Both resonances  have been recently experimentally confirmed  \cite{scalar}, but they remain slightly controversial.  

In order to extract the information about tetraquark states,  lattice QCD simulations evaluate correlation functions with tetraquark interpolators at the  source and the sink. 
In addition to possible tetraquarks, also the scattering states $P_1P_2$ ($\pi\pi$ for $I=0,2$ and $K\pi$ for $I=1/2,~3/2$) unavoidably contribute to the correlation function and this presents the main obstacle in extracting the information about tetraquarks. The scattering states $P_1(k)P_2(-k)$ at total momentum $\vec p=\vec 0$ have discrete energy levels  
\begin{equation}
\label{enP}
E_{P_1P_2}\simeq E_{P_1}(k)+E_{P_2}(-k)~,\quad {\mathrm{with}}\quad E_P(k)=\sqrt{m_P^2+\vec k^2}\quad {\mathrm{and }}\quad \vec k=\tfrac{2\pi}{L}\vec n
\end{equation}    
in the non-interacting approximation. The  energy level $E_{P_1P_2}\simeq m_{P_1}+m_{P_2}$ is low and makes an important contribution to the correlation functions. In order to identify possible tetraquarks, one has  to extract several energy levels $E_n$ in each isospin channel and then consider various  criteria that could distinguish between the one-particle (tetraquark) states  and the two-particle (scattering) states.
 We do not consider the more challenging $I=1$ channel, since there are two towers $K^+(k)\bar K^0(-k)$ and $\pi(k)\eta(-k)$ of scattering states.

The lattice  simulations \cite{mathur_scalar,tetra_sasa,tetra_jaffe,tetra_lat}
have not yet provided the final answer to whether the lightest scalar mesons are tetraquarks or conventional $\bar qq$  mesons. All previous tetraquark simulations were quenched and they ignored disconnected contractions (cf. Fig.~\ref{fig_contractions}). 
All simulations (except for \cite{tetra_sasa,tetra_jaffe}) consider only the $I=0$ channel and the simulations \cite{tetra_jaffe,tetra_lat} extract only the ground state. 
The strongest indication  for $\sigma$ as a tetraquark was obtained for $m_\pi\simeq 180-300$ MeV in \cite{mathur_scalar} by considering the lowest three energy levels\footnote{The $I=0$ energy levels   from \cite{mathur_scalar} and the present work are compared in the Section 3.3.}
 and the volume-dependence of the spectral weights. This impressive result on $\sigma$ meson  was obtained from a single correlator using the sequential empirical Bayes method \cite{mathur_scalar} and needs confirmation using a different method, for example the variational method used here. The first study that used the variational method to extract the ground and the excited energy levels in $I=0,1/2$ channels was presented in \cite{tetra_sasa}, but the first excited state was found much higher than $E_{P_1}(\tfrac{2\pi}{L})+ E_{P_2}(-\tfrac{2\pi}{L})$. The reason for that was attributed to the unfortunate choice of the interpolators that had the same color and Dirac  structure, while they differed only in  spatial structure. For this reason we take interpolators with different color and Dirac  structures in the present analysis, which enables us to extract the state $P_1(\tfrac{2\pi}{L})P_2(-\tfrac{2\pi}{L})$.  We note that there have been  few lattice simulation of tetraquarks or mesonic molecules in the related hidden charm sector \cite{XYZ_lattice}. 

In this paper we determine a spectrum of states with $J^{PC}=0^{++}$, $\vec p=\vec 0$ and $I=0,~2,~1/2,~3/2$ on the lattice using the variational method with a number of  tetraquark sources and sinks. We also determine  the couplings $\langle 0|{\cal O}_i|n\rangle$ between the interpolators ${\cal O}_i$ and the physical states $|n\rangle$. This is the first dynamical simulation intended to look for tetraquarks and we also perform  the quenched simulation in order to see whether there are any qualitative differences between the two cases. Our dynamical simulation has two flavors of Chirally Improved quarks, while our quenched simulation uses overlap quarks. In this pioneering study, we are interested in the ``pure'' tetraquark states with four valence quarks $\bar q\bar qqq$ and we prevent $\bar q\bar qqq\leftrightarrow \bar qq\leftrightarrow vac\leftrightarrow glue$ mixing by neglecting the disconnected contractions in $I=0,~1/2$ channels, as in all previous tetraquark studies\footnote{The disconnected contractions have been recently taken into account in the study of the ground scattering states with $I=0$ \cite{disconnected_zero} and $I=1/2$ \cite{disconnected_half}.}. 

The accurate lattice spectrum $E_n$ as a function of the lattice size $L$ in principle allows determination of the resonance mass and widths \cite{luscher,vol_dep}. The resonance appears as a state in the spectrum in addition to the discrete tower of scattering states. At the values of $m_\pi L$,  where the non-interacting energies of the resonance and the scattering state  would cross, the energy levels experience the largest energy shifts from the non-interacting values \cite{luscher,vol_dep}. These energy shifts in principle allow the determination of the resonance width. 
 In practice, the accurate determination of the excited energy levels on the lattice is challenging and only the $\rho$ meson width has 
 been  reliably determined from the ground energy level in this way \cite{width}. So far only  simulations of the toy models were able to  
extract the scattering states and the width of the  resonance 
 from the ground and  the  excited energy levels  \cite{toy_models}.

\begin{figure}[h!]
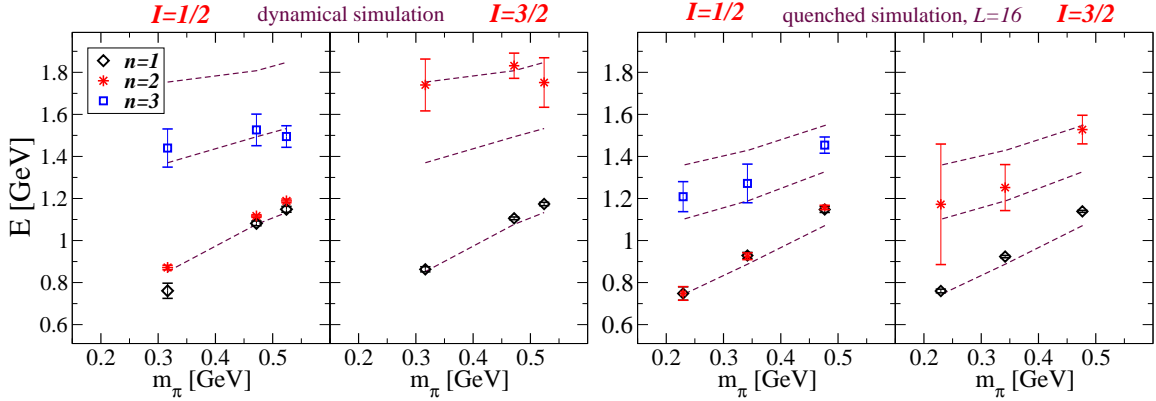

\begin{center}
\includegraphics[height=5.3cm,clip]{fig_dyn/dyn_spect_half_threehalf.eps}
\includegraphics[height=5.3cm,clip]{fig_q/kent_spect_half_threehalf.eps}
\end{center}
\caption{ \small The resulting spectrum $E_{n}$ for $I=1/2,~3/2$ in the dynamical and the quenched simulations. Note that there are two states ($n=1$ and 2) close to each other in $I=1/2$ case. The lines present the energies of non-interacting $K(k)\pi(-k)$ with $k=N~2\pi/L$ and $N=0,1,\sqrt{2}$. The plotted data is given in Tables \ref{tab_results_dyn} and \ref{tab_results_q}. }\label{fig_spect_half_threehalf}
\end{figure}

In the present work, our  excited energy levels are not 
accurate enough to allow for the determination of the $\sigma$ and 
$\kappa$ widths. We concentrate on a simpler and more realistic question: 
is there any  light state in addition to $P_1(0)P_2(0)$ and $P_1(\tfrac{2\pi}{L})P_2(-\tfrac{2\pi}{L})$ in the attractive channels $I=0$ or $I=1/2$?   Such an additional state could be related to $\sigma$ or $\kappa$ with sizable tetraquark components. Our main result for the spectrum, shown in Figs.~\ref{fig_spect_zero_two} and  \ref{fig_spect_half_threehalf}, indeed shows an additional light state  and we propose a possible interpretation that it is a tetraquark state. The extracted energy of the additional state  as a function of $m_\pi$ qualitatively agrees with $m_{\sigma,\kappa}(m_\pi)$ from unitarized Chiral Perturbation Theory (ChPT) \cite{pelaez}. 

The $\sigma$ and $\kappa$ are expected to become bound states at our heavier $m_\pi$, where decays $\sigma\to \pi\pi$ and $\kappa \to K\pi$ are no longer allowed kinematically \cite{pelaez}. In the range of $m_\pi$, where this might  occur,   we find  candidates for $\sigma$ and $\kappa$  close to the threshold,  in qualitative agreement with the prediction  of unitarized ChPT \cite{pelaez} and a lattice study of a toy-model with bound and scattering states \cite{sasaki}.

We also determine the spectra in repulsive channels $I=2$ and $I=3/2$, where no light resonance  is experimentally observed. Our main purpose here is to verify that there is no light state in addition to $P_1(0)P_2(0)$ and $P_1(\tfrac{2\pi}{L})P_2(-\tfrac{2\pi}{L})$. The results in Figs.~\ref{fig_spect_zero_two} and  \ref{fig_spect_half_threehalf}  demonstrate that indeed we do not find any additional state.

We explore two methods to distinguish the one-particle and two-particle states. The first is based on the time-dependence of the correlation functions and the eigenvalues of the variational method, which are explored analytically as well.  The second method is based on the volume dependence of the couplings  $\langle 0|{\cal O}_i|n\rangle$, which is explored in the quenched simulation and compared to the theoretical expectations.  

Some of our initial exploratory results have been  published in   proceedings  \cite{tetra_dyn_proceedings}. 

\vspace{0.2cm}

We present the methods to extract the spectrum and the $\langle 0|{\cal O}_i|n\rangle$ couplings in Sect. 2. 
The analytical expectations for the
time-dependence of the correlators and the 
eigenvalues at finite temporal extent are 
also given in this section, while certain derivations are delegated to the 
Appendix A.  Our numerical results and their  interpretation are given in  Sect. 3, and we end with  conclusions. In Appendix B we show that omission of the disconnected contractions cannot lead to an unphysical intermediate state with different isospin.

\begin{figure}[h!]
\begin{center}
\includegraphics[height=5.3cm,clip]{fig_dyn/dyn_spect_zero_two.eps}
\includegraphics[height=5.3cm,clip]{fig_q/kent_spect_zero_two.eps}
\end{center}
\caption{ \small The resulting spectrum $E_{n}$ for $I=0,~2$ in the dynamical (left) and the quenched (right) simulations. Note that there are two states ($n=1$ and 2) close to each other in the $I=0$ case. The lines present the energies of non-interacting $\pi(k)\pi(-k)$ with $k=N~2\pi/L$ and $N=0,1,\sqrt{2}$. The plotted data is given in Tables \ref{tab_results_dyn} and \ref{tab_results_q}.}\label{fig_spect_zero_two}
\end{figure}

\section{Simulation}

\subsection{Correlation matrix and interpolators}

We computed the same tetraquark correlation functions $C_{ij}(t)$  in a two-flavor dynamical simulation and in a quenched simulation. The purpose was   to see whether there are any significant qualitative differences between the two cases. The results, presented in the Section \ref{sec_results}, show qualitative agreement between dynamical and quenched results.  
The details about both simulation are given in Section \ref{sec_simulation}.
 
The tetraquark correlation function creates  a state  $\bar q\bar qqq$ with chosen $I$ and $J^{PC}=0^{++}$    at $t=0$ and annihilates it at some later time $t$, where the projection to the total momentum $\vec p=\vec 0$ is made. 
The time-dependence  
 is obtained by inserting a complete set of physical states $|n\rangle$ with given quantum numbers  
\begin{equation}
\label{cor}
C_{ij}(t)=\langle 0| {\cal O}_i(t){\cal O}^{\dagger}_j(0)|0\rangle_{\vec p =\vec 0}=\sum_{\vec x}\langle 0| {\cal O}_i(\vec x,t){\cal O}^{\dagger}_j(\vec 0,0)|0\rangle\stackrel{T\to \infty}{\longrightarrow} \sum_n Z_i^{n}Z_j^{n*}~e^{-E_n~t}~\quad n=1,2,\cdots
\end{equation} 
with coupling $Z_i^{n}\equiv\langle 0| {\cal O}_i|n\rangle$ and temporal extent of the lattice $T$.  The correlation matrix is used to extract energy levels $E_n$ and couplings $Z_i^{n}$ for the tetraquark system. 

We consider five interpolators in the case of $I=0,~1/2$ and three interpolators in the case of $I=2,~3/2$. The interpolators differ only in Dirac and color structure, but they have the same spatial structure: all quark fields are evaluated at the same space-time point.  

For $I\!=\!0$ and $1/2$, where a resonances may be expected, the first three interpolators  ${\cal O}_{i=1,2,3}$ are products of two color-singlet currents  
 (with sum over spatial components $\mu$ for  ${\cal O}_{i=2,3}$) and the flavor structure is derived in Appendix B. 
The last two interpolators are of well-known diquark anti-diquark type \cite{tetra_phenomenology}
\begin{align}
\label{interpolators_zero_half}
{\cal O}^{I=0}_{i=1,2,3}&=\sum_{\mu=1,2,3}2(\bar d\Gamma_i^\mu u)(\bar u\Gamma_i^\mu d)+\tfrac{1}{2}(\bar u\Gamma_i^\mu u)(\bar u\Gamma_i^\mu u)+\tfrac{1}{2}(\bar d\Gamma_i^\mu d)(\bar d\Gamma_i^\mu d)-(\bar u\Gamma_i^\mu u)(\bar d\Gamma_i^\mu d)\;,\nonumber\\
{\cal O}^{I=0}_{i=4,5}&=[\bar u \bar \Gamma_i \bar d^T]_a~[ u^T  \Gamma_i d]_a\;,\nonumber\\ 
{\cal O}^{I=1/2}_{i=1,2,3}&=\sum_{\mu=1,2,3}\sum_{q=u,d,s} (\bar s\Gamma_i^\mu q)(\bar q\Gamma_i^\mu u)\;,\nonumber\\
{\cal O}^{I=1/2}_{i=4,5}&=[\bar s \bar \Gamma_i \bar d^T]_a~[ u^T \Gamma_i d]_a\;.
\end{align}
 Here $\bar \Gamma\equiv \gamma_0 \Gamma^\dagger \gamma_0$  and
\begin{equation}
\label{Gamma}
\Gamma_1=\gamma_5~,\ \Gamma_2^\mu=\gamma^\mu~,\ \Gamma_3^\mu=\gamma^\mu\gamma_5~,\ \Gamma_4=C\gamma_5~,\ \Gamma_5=C~,
\end{equation}
 while $[q^T\Gamma Q]$ denotes a (pseudo) scalar diquark 
$[q^T\Gamma Q]_a\equiv \epsilon _{abc} [q_b^T \Gamma Q_c-Q_b^T\Gamma q_c]$.
The  $I=1/2$ tetraquark interpolators above are constructed to transform as $|I,I_3\rangle=|1/2,1/2\rangle$ under $SU(2)_{F}$ and like $\bar su$ flavor state under $SU(3)_{F}$.

We use a smaller three-dimensional interpolator basis for $I\!=\!2$ and $3/2$, which are not of our prime interest since no resonances have been observed  experimentally in these repulsive channels.  All these interpolators are of current-current type
 \begin{align}
\label{interpolators_two_threehalf} 
{\cal O}^{I=2}_{i=1,2,3}&=\sum_{\mu=1,2,3} (\bar d\Gamma_i^\mu u)(\bar d\Gamma_i^\mu u)\nonumber\\
{\cal O}^{I=3/2}_{i=1,2,3}&=\sum_{\mu=1,2,3} (\bar s\Gamma_i^\mu u)(\bar d\Gamma_i^\mu u)
\end{align}
with $\Gamma_i$ as defined in  (\ref{Gamma}). 

\begin{figure}[t!]
\begin{center}
\includegraphics[height=2.5cm,clip]{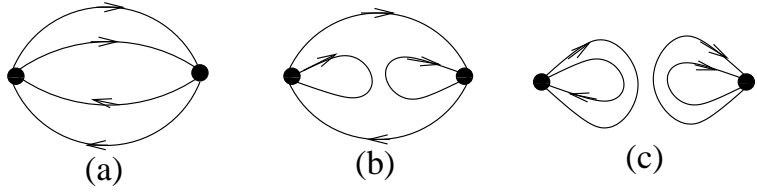}
\end{center}
\caption{ \small Quark contractions for our tetraquark correlators: connected (a), singly disconnected (b) and doubly disconnected (c). Only connected contractions are  taken into account in our simulation, for reasons explained in the  text.   }\label{fig_contractions}
\end{figure}

Fig.~\ref{fig_contractions} shows the contractions which enter the correlation matrix with  our tetraquark interpolators at the source and the sink: $I=0$ has all three contractions, $I=1/2$ has contractions (a,b) and $I=2,3/2$ have  only the connected contraction (a). In this pioneering study, we are interested in physical states with four valence quarks $\bar q\bar qqq$ (``pure'' tetraquark states and $P_1P_2$ states) and we therefore take into account only the connected contractions. The singly (doubly) disconnected contraction couples also to $\bar qq$ (vacuum and glueball) states  and we ignore them in order to be able to attribute a definite valence $\bar q\bar qqq$ quark structure to the obtained physical states. Another reason for omitting the  disconnected contractions is that they are difficult to evaluate and they are often noisy. 
We note that  it is not legitimate to  ignore the 
disconnected contractions in a proper field theory, as it leads to the violation of the unitarity.  A possible effect of this approximation on our results is discussed in Section  \ref{sec_interpretation} and in Appendix B. 
We choose this approximation in our study with the given physical motivation and leave a proper study of physical states including mixing $\bar q\bar qqq\leftrightarrow \bar qq\leftrightarrow vac\leftrightarrow glue$ for the future.

\subsection{Variational method at $T\to \infty$}

The extraction of the energies $E_n$ and the couplings $Z_i^{n}$ from the correlation functions (\ref{cor}) using multi-exponential fits is unstable. Instead, we use our $N\times N$ correlation matrix to compute the eigenvalues $\lambda^{n}(t)$ and eigenvectors $\vec u^{n}(t)$ of the generalized eigenvalues problem \cite{var_luscher,var_alpha}
\begin{equation}
C(t)~\vec u^{n}(t)=\lambda^{n}(t,t_0)~ C(t_0)~\vec u^{n}(t) ~.
\end{equation}
The energy  can be extracted from 
\begin{equation}
\label{lam}
\lambda^{n}(t)\stackrel{t\to \infty}{\longrightarrow}e^{-E_n(t-t_0)}~
\end{equation}
when the temporal extent of the lattice $T$ is very large. 
The  error on the extracted energy $E_n$ due to finite basis $N$ is ${\cal O}(e^{-(E_{N+1}-E_n)t})$ for $t_0\leq t \leq 2t_0$ \cite{var_alpha}. We will demonstrate 
that our results are almost independent of $t_0$ for $t_0\in[1,4]$, while they get noisier for $t_0\geq 5$. Our main analysis is based on $t\in[6,10]$, so the condition $t\leq 2t_0$ is  satisfied or close to being satisfied. The exponential time-dependence (\ref{lam}) would apply only for infinite temporal extent $T$ and we discuss the significant effect of finite 
$T$  on $C_{ij}(t)$ and $\lambda^{n}(t)$ in Sect. \ref{sec_finite_T}.   

The eigenvectors $\vec u^{n}(t)$, which satisfy the orthogonality relation $(\vec u^{n},C(t)\vec u^{(m)})\propto \delta_{nm}$,  
allow us to determine the couplings $Z_i^{n}$  at large $t$ (see for example the derivation in \cite{lang_rho}) 
\begin{equation}
\label{Z1}
|Z_i^{n}|= |\langle 0|{\cal O}_i|n\rangle|=\frac{|\sum_k C_{ik}(t)~ u_k^{n}(t)|}{\sqrt{ \sum_{lm} |u_l^{n*}(t) ~C_{lm}(t)~u_m^{n}(t)|}}~e^{E_n t/2}~.
\end{equation}
Note that the normalization of $\vec u^{n}(t)$, which is arbitrary, cancels in (\ref{Z1}). The error on the extracted coupling  $Z_i^{n}$ due to the finite basis $N$ is ${\cal O}(e^{-(E_{N+1}-E_n)t_0})$ for $t\leq 2\ t_0$ and for fixed $t-t_0$ \cite{var_alpha}. Extracting the ratio of couplings for a given state $|n\rangle $ to two different interpolators is particularly straightforward
\begin{equation}
\label{rat_Z}
\biggl\vert \frac{Z_i^{n}}{Z_j^{n}}\biggr\vert=\frac{|\sum_k C_{ik}(t)~ u_k^{n}(t)|}{|\sum_{k^\prime} C_{jk^\prime}(t) u_{k^\prime}^{n}(t)|}~.
\end{equation}
Extracting $|Z_i^{n}|$ itself 
\begin{equation}
\label{Z}
\frac{|\sum_k C_{ik}(t)~ u_k^{n}(t)|}{\sqrt{ \sum_{lm} |u_l^{n*}(t) ~C_{lm}(t)~u_m^{n}(t)|}}=|Z_i^{n}|~e^{-B t}
\end{equation}
 requires fitting the LHS to the form on the RHS. 
 We always verify that the fitted coefficient $B$ in the exponent is consistent with $E_n/2$ as obtained from $\lambda^{n}(t)$. \\

\subsection{ Effect of finite $T$ on the correlation matrix and the eigenvalues}
\label{sec_finite_T}

Our temporal extents ($T=32$ for the dynamical and $T=28$ for the quenched simulation) are not very large, so we need to understand the effect of finite $T$ on $C_{ij}(t)$ and $\lambda^n(t)$.  
At finite temporal extent $T$, the time-dependence $e^{-E_n t}$ gets modified 
depending on the boundary conditions and the  nature of the states. 
We use anti-periodic boundary conditions in the time direction for quarks and anti-quarks.
 
If a single one-particle state $|n\rangle$ dominates the correlator, the diagonal correlator $C_{ii}(t)$ behaves as 
\begin{equation}
\label{time_cosh}
C_{ii}(t)\stackrel{{\mathrm{large}}\ t}{\longrightarrow} |Z_i^{n}|^2~[e^{-E_nt}+e^{-E_n(T-t)}]~. 
\end{equation}
If the correlator is dominated by a two-particle state $|n\rangle=|P_1P_2\rangle $, it behaves as (see  Appendix A of this paper and Appendix A of \cite{tetra_sasa} and \cite{savage})
\begin{equation}
\label{time_scatter}
C_{ii}(t) \stackrel{{\mathrm{large}}\ t}{\longrightarrow} |Z_i^{n}|^2~[e^{-E_nt}+e^{-E_n(T-t)}]+|\tilde Z_i^{n}|^2~[e^{-m_{P_1}t}e^{-m_{P_2}(T-t)}+ e^{-m_{P_2}t}e^{-m_{P_1}(T-t)}]~, 
\end{equation}
where $E_n$ is two-particle energy. 
Let us consider the 
 relative importance of the couplings $Z_i^{n}=\langle 0|{\cal O}_i|P_1P_2\rangle$ and $\tilde Z_i^{n} =\langle P_1^\dagger |{\cal O}_i|P_2\rangle$ \cite{tetra_sasa}, which  will be needed in our further study.  
Both matrix elements have similar structure, therefore one expects that $Z_i^{n}$ is of the same order of magnitude as $\tilde Z_i^{n}$. However we do {\it not} expect that $\langle P_1^\dagger|{\cal O}_i|P_2\rangle$ is exactly equal to $\langle 0|{\cal O}_i|P_1P_2\rangle$: in the second case ${\cal O}_i$ annihilates the (interacting) state $P_1P_2$, where $P_1$ and $P_2$ existed at the same time and therefore interacted; in the first case   ${\cal O}_i$ annihilates the $P_2$
and creates $P_1^\dagger$, so $P_1$ and $P_2$ never exist at the same time and there is no interaction between them. We therefore believe that in the interacting theory $ \langle P_1^\dagger|{\cal O}_i|P_2\rangle\not = \langle 0|{\cal O}_i|P_1P_2\rangle$ and $Z_i^{n}\not = \tilde Z_i^{n}$,  but we expect they are of the same order of magnitude.

In reality, several physical states contribute to the correlation matrix and the time-dependence of the eigenvalues becomes more complicated. We are in particular interested in the cases where  two-particle states and also possible one-particle (tetraquark) states  contribute to the correlation matrix. We study the  generalized eigenvalue problem for such a situation in the Appendix A. We find that  
\begin{enumerate}
\item[(i)] the eigenvalue corresponding to the one-particle state  would have a time-dependence proportional to $e^{-E_nt}+e^{-E_n(T-t)}$ and
\item[(ii)] the  eigenvalue corresponding to the two-particle state would have a time-dependence proportional to  $e^{-E_nt}+e^{-E_n(T-t)}+R[e^{-m_{P_1}t}e^{-m_{P_2}(T-t)}+ e^{-m_{P_2}t}e^{-m_{P_1}(T-t)}]$ with two-particle energy $E_n$ 
\end{enumerate}
only if the following relations would apply exactly: $ \langle P_1^\dagger|{\cal O}_i|P_2\rangle= \langle 0|{\cal O}_i|P_1P_2\rangle$ or more generally if $\langle P_1^\dagger|{\cal O}_i|P_2\rangle= R ~\langle 0|{\cal O}_i|P_1P_2\rangle$ with $R$ independent of $i$. However, we argued in the previous paragraph that these relations do {\it not} apply exactly and so the time-dependence of eigenvalues is more complicated than in statements (i,ii) above.  In the Appendix A we  argue (although this has not been rigorously proved) that  the two-particle as well as the one-particle eigenvalues have the general form\footnote{This form applies if only one two-particle state $P_1P_2$ makes significant contribution at $t\simeq T/2$. }
\begin{equation}
\label{time_lam}
\lambda^{n}(t) = w^n~[e^{-E_nt}+e^{-E_n(T-t)}]+\tilde w^n ~[e^{-m_{P_1}t}e^{-m_{P_2}(T-t)}+ e^{-m_{P_2}t}e^{-m_{P_1}(T-t)}]~. 
\end{equation}
In our analysis, we extract $E_n$ from eigenvalues $\lambda^{n}(t)$ using a three-parameter fit ($E_n,~w,~\tilde w$), where masses of $P_{1,2}=\pi,~K$ are fixed\footnote{We verified that  the variation of the results is negligible if $m_{\pi,K}$ are varied in the ranges given in Tables \ref{tab_run_dyn} and \ref{tab_run_q}.} to the measured values given in Tables \ref{tab_run_dyn} and \ref{tab_run_q}.   In case of equality $ \langle P_1^\dagger|{\cal O}_i|P_2\rangle= \langle 0|{\cal O}_i|P_1P_2\rangle$,  we would expect $\tilde w=0$ for one-particle state and $\tilde w=w$ for two-particle state. We note that the extracted $w^n$ and $\tilde w^n$ in Tables \ref{tab_results_dyn} and \ref{tab_results_q}  are  of similar magnitude.  

Finally we note that the form (\ref{time_lam}) generally applies only in case when all diagonal and non-diagonal correlators are symmetric with respect to 
$t \to T-t$  and that all our $C_{ij}(t)$ have this property. In case when some parts of non-diagonal correlators are anti-symmetric with respect to $t\leftrightarrow T-t$, ``backward propagation'' described in Section IIF of \cite{ci_derivative} may appear.

\subsection{Details of dynamical and quenched lattice simulations}\label{sec_simulation}\label{sec_T}

Our correlation functions are constructed based on gauge configurations and
quark propagators for two cases:  the two-flavor dynamical simulation  \cite{ci_dyn} and  the  quenched simulation \cite{mathur_scalar}. 

\begin{itemize}
\item Details of the {\bf two-flavor dynamical simulation} together with  the ground state hadron spectroscopy are given  in \cite{ci_dyn}, while some related results based on a similar setup are given in \cite{ci_derivative,ci_references}.   In that simulation the  L\"uscher-Weisz gauge action \cite{luscher_weisz}
and  two-flavors of dynamical degenerate Chirally Improved (CI) quarks \cite{ci_original} have been used.
The gauge fields are periodic in all four space-time directions, the fermion field anti-periodic in the time direction.
One level of stout smearing \cite{stout} is applied to the gauge configurations, which is  considered as a part of full Dirac operator.   
We use three ensembles (C,B,A) with the same lattice volume $16^3\times 32$ and three different $u/d$ quark masses, corresponding to $m_\pi\simeq 318-526$ MeV (see Table \ref{tab_run_dyn}).  The lattice spacings, also given in Table \ref{tab_run_dyn}, have been determined using $r_0= 0.48$ fm and are close to $a\simeq 0.15$ fm for all three ensembles.  

The valence $u,\,d,\,s$ quarks are also of Chirally Improved type. 
The valence $u/d$ quark masses are always fixed to the dynamical $u/d$ quark masses in our study. The valence strange quark masses is fixed  from $m_\Omega$. 

All quark sources and sinks are Jacobi-smeared by applying 
\begin{equation}
\sum_{n=0}^{N}(\kappa~H)^n\qquad {\mathrm {with}} \qquad H=\sum_{j=1}^3\bigl[U_j(\vec x,t)\delta_{\vec x+\vec j,\vec y}+U^\dagger_j(\vec x-\vec j,t)\delta_{\vec x-\vec j,\vec y}\bigr]~, 
 \end{equation}
which is invariant under rotations and preserves the interpolator quantum numbers. For this analysis we use a single (``narrow'') smearing with the values of $N$ and $\kappa$ in Table \ref{tab_run_dyn}, which are chosen to give the source/sink Gaussian width of approximately $0.27$ fm.

\item Details of the {\bf quenched simulation} are presented in \cite{mathur_scalar,chiral_logs}. It employs overlap valence quarks, which have exact chiral symmetry even at finite lattice spacing. The gauge fields are generated using Iwasaki actions, where the lattice spacing $a=0.200(3)$ fm is determined using $f_\pi$. Our main analysis is based on the volume $16^3\times 28$, while a smaller volume $12^3\times 28$ with the same lattice spacing is used for the study of the volume-dependence of couplings $\langle 0|{\cal O}_i|n\rangle$.   
All $u,d,s$ quarks have point-like sources and sinks, while the strange quark mass is fixed from $m_\phi$. We use three $u/d$ quark masses, corresponding to $m_\pi=230-478$ MeV (see Table \ref{tab_run_q}).  
   
\end{itemize}

\begin{figure}[t!]
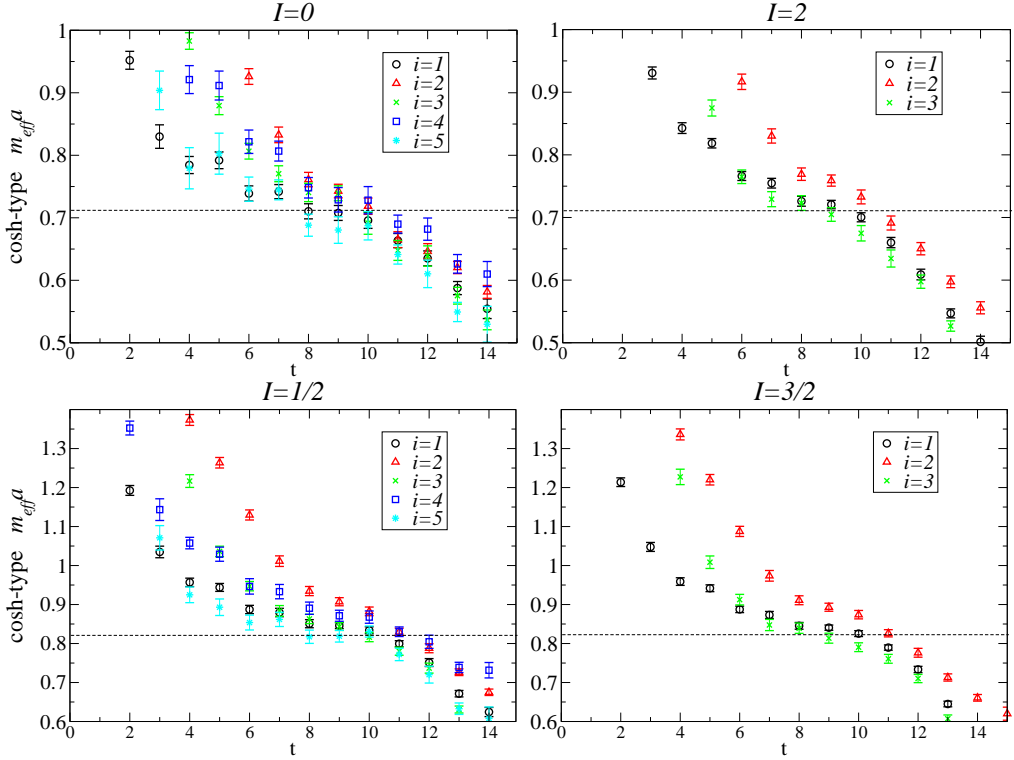

\begin{center}
\includegraphics[height=5cm,clip]{fig_dyn/eff_c_diag_zero_B.eps}
\includegraphics[height=5cm,clip]{fig_dyn/eff_c_diag_two_B.eps}
\includegraphics[height=5cm,clip]{fig_dyn/eff_c_diag_half_B.eps}
\includegraphics[height=5cm,clip]{fig_dyn/eff_c_diag_threehalf_B.eps}
\end{center}
\caption{ \small Cosh-type effective mass (\ref{eff_cosh}) for the diagonal  correlators $C_{ii}(t)$  
at $m_\pi=469$ MeV in the dynamical simulation. The lines indicate $2m_\pi$ or $m_\pi+m_K$.  }\label{fig_c_diag}
\end{figure}

\section{Results}\label{sec_results}

\subsection{Time-dependence of diagonal correlators}

The effective masses for  diagonal correlators $C_{ii}(t)$ for four isospins 
are displayed in Fig.~\ref{fig_c_diag}. We show the 
 {\it cosh-type effective 
mass}, defined as 
\begin{equation}
\label{eff_cosh}
\frac{F(t)}{F(t+1)}=\frac{e^{-m_{eff}^{t}~t}+e^{-m_{eff}^{t}~(T-t)}}{e^{-m_{eff}^{t}~(t+1)}+e^{-m_{eff}^{t}~(T-t-1)}}~,\quad F(t)=C_{ii}(t)\  {\mathrm{or}}\ \lambda^n(t)
\end{equation}
so that $m_{eff}$ equals the energy $E$ if 
$F(t)=w[ e^{-Et}+e^{-E(T-t)}]$. The observed effective masses have 
 sizable excited state contributions at small $t$ and they have  significant 
drop for $t>10$, which indicates that a two-particle state $|n\rangle$
dominates $C_{ii}(t)$ at $t\simeq T/2$ 
for each $i$ and $I$ (see Sect. \ref{sec_T}).

\begin{figure}[t!]
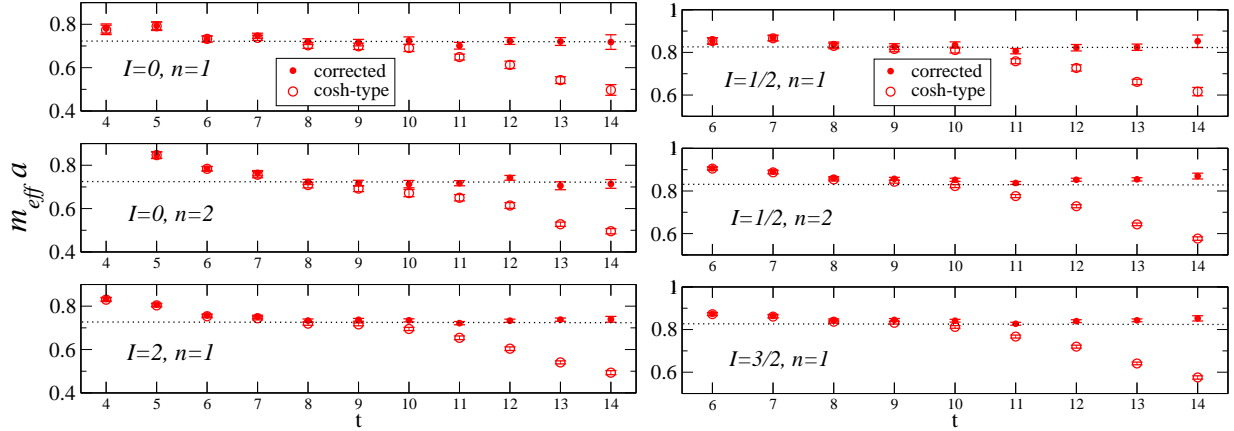

\begin{center}
\includegraphics[height=5.7cm,clip]{fig_dyn/effcorrected_zero_two.eps}
\includegraphics[height=5.7cm,clip]{fig_dyn/effcorrected_half_threehalf.eps}
\end{center}
\caption{ \small The typical  cosh-type effective masses according to (\ref{eff_cosh}) (open symbols) and the corrected effective masses (\ref{eff_corrected}) (filled symbols) for two lightest states in $I=0,~1/2$ channels and for the lightest state in $I=2,~3/2$ channels. The full $5\times 5$ correlation matrix is used for $I=0,~1/2$ and the $3\times 3$ correlation matrix is used for $I=2,~3/2$. The figure shows as an example our results for the dynamical simulation with $m_\pi=469$ MeV and  $t_0=3$, with $\tilde w$ fitted via (\ref{time_lam})  in the range $t\in [8,15]$. The lines represent $2m_\pi$ ($I=0,~2$) and $m_\pi+m_K$ ($I=1/2,~3/2$).   }\label{fig_effcorrected}
\end{figure}

\subsection{Time-dependence of eigenvalues}

Due to the lack of 
reliable plateaus in the diagonal elements of the correlation function 
we turn to use the eigenvalues of the variational method in order
to extract the spectrum. 
Typical cosh-type effective masses (\ref{eff_cosh}) for eigenvalues   
are plotted by empty symbols in Fig.~\ref{fig_effcorrected}. 
The excited state contribution at small $t$ 
is smaller than for the diagonal correlators and $m_{eff}$  
reach a short plateau, 
but then they drop significantly  for $t>10$.  
We note that the
 drop is less significant for smaller $m_\pi$ and more significant for 
larger $m_\pi$, as illustrated in \cite{tetra_sasa}. 
The drop demonstrates  that eigenvalues don't have a simple $e^{-E_nt}+e^{-E_n(T-t)}$ time-dependence, but have a more complicated time-dependence (\ref{time_lam}) due to the presence of two-particle states $\pi\pi$ or $K\pi$ in the box with finite $T$. 
The  {\it corrected effective mass}  takes that effect into account  
\begin{equation}
\label{eff_corrected}
\frac{F(t)- \tilde w [e^{-m_{P_1} t}e^{-m_{P_2} (T-t)} + \{t\!\leftrightarrow\! T\!-\!t\}]}{F(t\!+\!1)-\! \tilde w [e^{-m_{P_1} (t+1)}e^{-m_{P_2} (T-t-1)} + \{t\!\leftrightarrow \!T\!-\!t\}]}=\frac{e^{-m_{eff}^{t}t}+e^{-m_{eff}^{t}(T-t)}}{e^{-m_{eff}^{t}(t+1)}+e^{-m_{eff}^{t}(T-t-1)}}~,
\end{equation}
  so $m_{eff}$ is equal to $E_n$ when $F(t)=C_{ii}(t)$ or  $F(t)=\lambda^n(t)$  has the form (\ref{time_lam}). Such $m_{eff}$  is obtained   after $\tilde w$ has been determined by fitting $\lambda^n(t)$ to  (\ref{time_lam}). The corrected effective mass
 is presented by the full symbols in Fig.~\ref{fig_effcorrected}, it is flat 
and  it demonstrates that eigenvalues really have time-dependence 
of the form (\ref{time_lam}).  

We will always use the fitting form (\ref{time_lam}) and the 
corrected effective mass  (\ref{eff_corrected}) 
for $\lambda_{I=0,1/2}^{n=1,2}$ and for $\lambda_{I=2,3/2}^{n=1}$. For higher states $\lambda_{I=0,1/2}^{n\geq 3}$ and $\lambda_{I=2,3/2}^{n\geq 2}$, the error-bars on $\lambda^n(t)$ are large at $t>10$, 
where the finite $T$ effect is significant,  
such that the three-parameter fit with $(E_n,w^n,\tilde w^n)$ is not stable. 
In this case we will present cosh-type effective masses (\ref{eff_cosh}) 
 and we will fit to 
\begin{equation}
\label{time_lam_naive}
\lambda^{n}(t) = w^n~[e^{-E_nt}+e^{-E_n(T-t)}]
\end{equation}
at rather small $t$, where the finite $T$ effect is not significant.  

Let us note that the significant effect at finite $T=28,~32$ prevents us from a reliable determination of the  energy shifts $\Delta E_n=E_n-m_{P_1}-m_{P_2}$, which require very long stable plateaus. Therefore, we do not aim at determining the energy shifts, but we determine the spectrum $E_{n=1,2,3}$ itself with a reasonable precision.
   
\begin{figure}[t!]
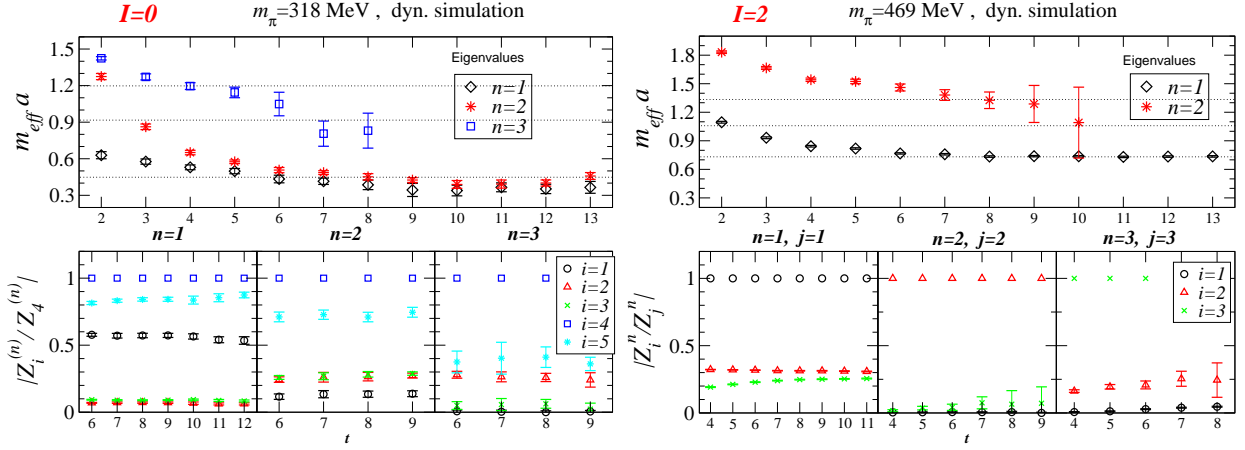

\begin{center}
\includegraphics[height=5.9cm,clip]{fig_dyn/dyn_zero_C.eps} 
\includegraphics[height=5.9cm,clip]{fig_dyn/dyn_two_B.eps}
\end{center}
\caption{ \small  Typical effective masses of the eigenvalues $\lambda^{n}(t)$ for $I=0,~2$. Corresponding ratios  $|Z_i^{n}/Z_j^{n}|$ at given $n$  are also shown ($j$ is the largest component). The full $5\times 5$ matrix is used for $I=0$ and the $3\times 3$ correlation matrix is used for $I=2$.  
Results for a specific  $u/d$  masses and $t_0=1$ in the dynamical simulation are shown. The  lines present the energies of non-interacting $\pi(k)\pi(-k)$ with $k=N 2\pi/L$ and $N=0,1,\sqrt{2}$. }\label{fig_eff_zero_two}
\end{figure} 

\subsection{Energy levels $E_n$ and couplings $Z_i^n$}

\subsubsection*{\bf $I=0$ and $I=2$}

The typical effective masses  for $I=0,2$ are collected in Fig.~\ref{fig_eff_zero_two}. The ratios of couplings $|Z_i^n/Z_j^n|$ extracted via (\ref{rat_Z}) are also shown. 
The lines display the three lowest energies of $\pi(k)\pi(-k)$ in the non-interacting case. 

\begin{figure}[t!]
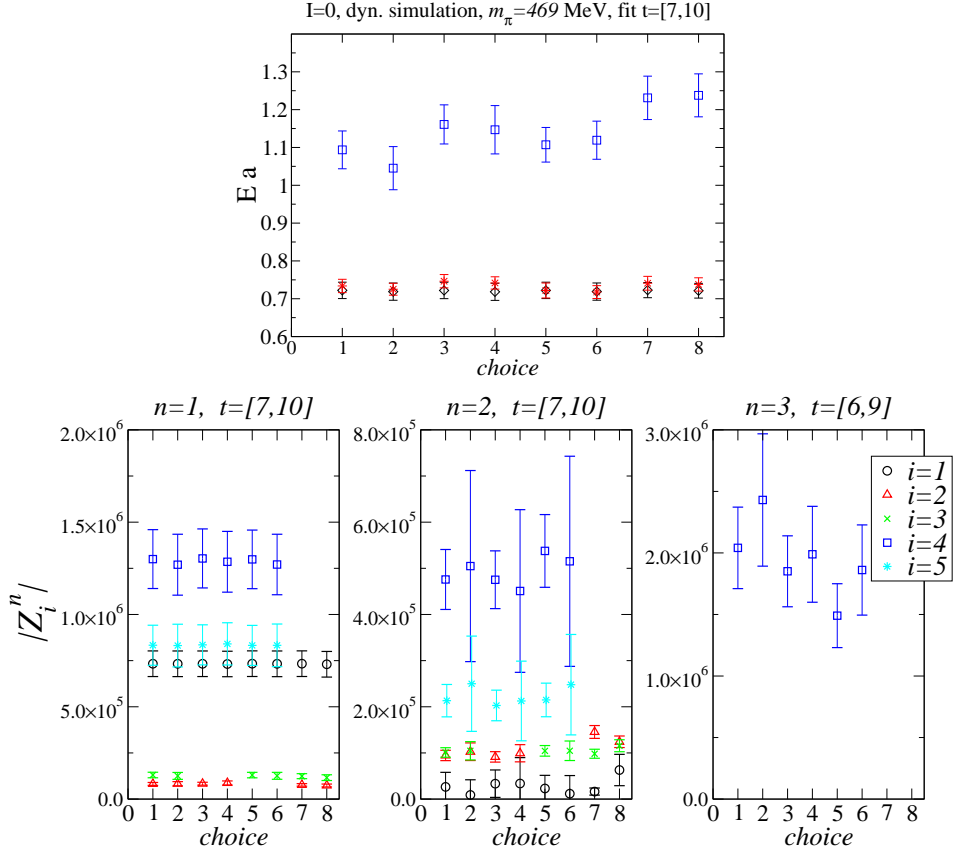

\begin{center}
\includegraphics[height=5cm,clip]{fig_dyn/m_dependence_I_zero_runB.eps}\vspace{6pt}\\
\includegraphics[height=6cm,clip]{fig_dyn/Z_dependence_I_zero_runB_1.eps} 
\includegraphics[height=6cm,clip]{fig_dyn/Z_dependence_I_zero_runB_2.eps}
\includegraphics[height=6cm,clip]{fig_dyn/Z_dependence_I_zero_runB_3.eps} 
\end{center}
\caption{ \small   Extracted energies  $E_n$ and couplings $Z_i^n=\langle 0| {\cal O}_i^n|n\rangle $ in the $I=0$ channel for various choices of $t_0$ and interpolator sets. A typical case with $m_\pi=469$ MeV in the dynamical simulation is shown. On the horizontal axis $choice=1,3,5,7$ correspond to sets ${\cal O}_{12345},~{\cal O}_{1245},~{\cal O}_{1345},~{\cal O}_{123}$, respectively, where $t_0=2$ is taken in  all these cases. Similarly, $choice=2,4,6,8$ correspond to the same four sets and $t_0=3$. The values $E_3$ from interpolator set ${\cal O}_{1,2,3}$ are naturally too high since a
$3\times 3$ matrix cannot reliably provide $E_3$.
  The $Z_i^{n=1,2}$  for the ground and the first excited states  are independent of the choice to a good precision. For the second excited state, we show only the largest component $Z_4^{n=3}$, which is expected to be the most reliable among all components.  }\label{fig_m_Z_dependence_zero}
\end{figure}

\begin{figure}[t!]
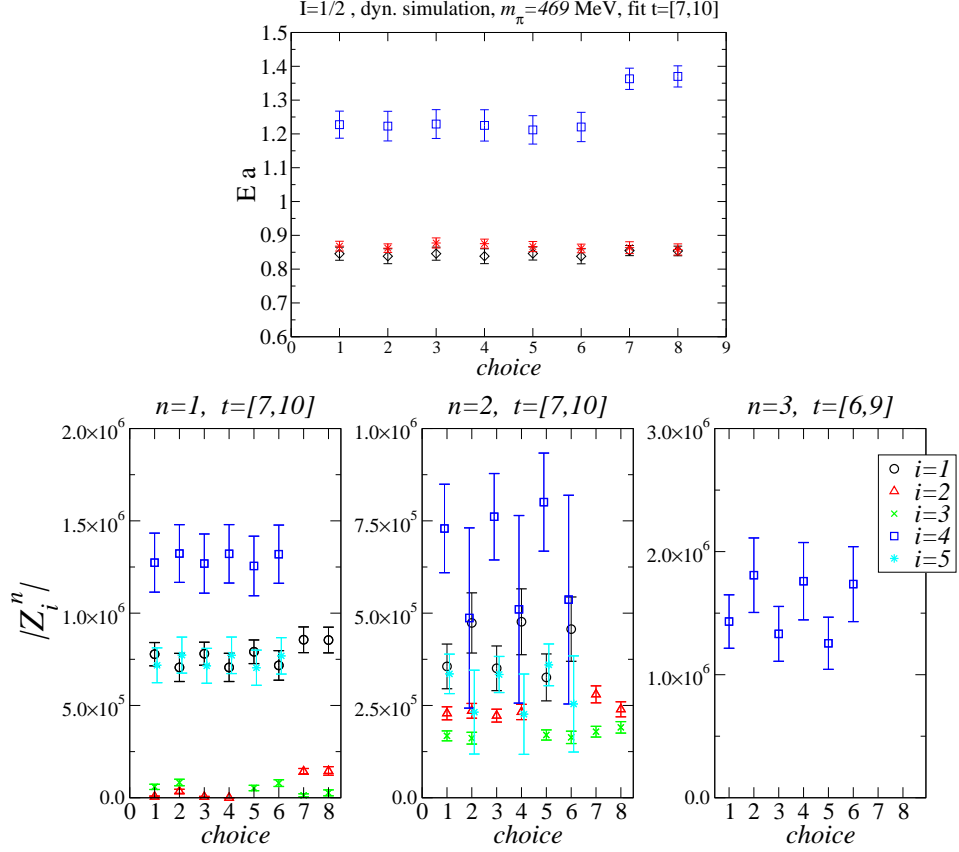

\begin{center}
\includegraphics[height=5cm,clip]{fig_dyn/m_dependence_I_half_runB.eps}\vspace{6pt}\\
\includegraphics[height=6cm,clip]{fig_dyn/Z_dependence_I_half_runB_1.eps}
\includegraphics[height=6cm,clip]{fig_dyn/Z_dependence_I_half_runB_2.eps}
\includegraphics[height=6cm,clip]{fig_dyn/Z_dependence_I_half_runB_3.eps}
\end{center}
\caption{ \small  Analogous to Fig.~\ref{fig_m_Z_dependence_zero}, but for $I=1/2$.  }\label{fig_m_Z_dependence_half}
\end{figure}

In $I=0$ case, we find one state with  energy close to $\pi(0)\pi(0)$, another state with energy close to $\pi(\tfrac{2\pi}{L})\pi(-\tfrac{2\pi}{L})$ and we also find an additional light state (close to the lowest state). 
This applies for all quark masses and for the dynamical as well as the quenched simulation. We varied  $t_0\in [1,4]$ and performed the diagonalization of $5\times 5$ and all possible $4\times 4$ and $3\times 3$ sub-matrices. We find that the extracted energies\footnote{As an exception, all energies in Figs.~\ref{fig_m_Z_dependence_zero} and \ref{fig_m_Z_dependence_half}  are extracted using the fit form (\ref{time_lam_naive}) 
at rather small $t\in[7,10]$, where a finite $T$ effect is not significant. The resulting $E_n$ are just intended to demonstrate independence on $t_0$ and on the choice of the interpolator set. }  $E_{n}$ and  couplings\footnote{The  $|Z_i^n|$ was determined via (\ref{Z}) by fitting in the time range indicated in the plot.} $|Z_i^n|$ are almost independent of these choices for all quark masses and both simulations, 
as demonstrated for a specific case in Fig.~\ref{fig_m_Z_dependence_zero}.

In the $I=2$ case, we find one state with energy close to $\pi(0)\pi(0)$, another state with energy close to $\pi(\tfrac{2\pi}{L})\pi(-\tfrac{2\pi}{L})$ and no additional light state (see Fig.~\ref{fig_eff_zero_two}). Again, this applies for all quark masses, both simulations and for the range of $t_0\in [1,4]$. In this case we use only a $3\times 3$ matrix (\ref{interpolators_two_threehalf}), which is probably not large enough to capture the energy of  $\pi(\tfrac{2\pi}{L})\pi(-\tfrac{2\pi}{L})$ exactly (it naturally comes out too high). We point out that our intention was not to capture the energy of $\pi(\tfrac{2\pi}{L})\pi(-\tfrac{2\pi}{L})$ correctly, but to verify that there is no light state in addition to $\pi(0)\pi(0)$ in the $I=2$ channel.

The final result for the dependence of the  extracted spectrum on $m_\pi$ in both simulations is presented in Fig.~\ref{fig_spect_zero_two}. Tables \ref{tab_results_dyn} and \ref{tab_results_q} provide the corresponding numerical results, together with the choices of $t_0$, interpolator sets and fit ranges. The values are obtained using an uncorrelated fit to (\ref{time_lam}) or  (\ref{time_lam_naive}) and the error-bars are obtained with the  single elimination jack-knife method.

Finally, we compare our $I=0$ results to those obtained from a single $\pi\pi$ 
correlator using the sequential empirical Bayes method (SEBM) \cite{mathur_scalar}.  Above $300$ MeV, we get degenerate results for the two states below $\pi(\tfrac{2\pi}{L})\pi(-\tfrac{2\pi}{L})$.  For $m_\pi > 300$ MeV, the authors of \cite{mathur_scalar} were unable to 
separate these two states using the SEBM; there is no disagreement since the 
SEBM is 
not designed to resolve degenerate states. Although the 
$m_\pi> 300$ MeV results 
were not included in the publication, the authors of \cite{mathur_scalar}
 do present results 
for pion masses in the range  $182$ MeV to $250$ MeV. Both groups have 
analyzed  a pion mass of $230$ MeV. For this, the authors of 
\cite{mathur_scalar} get the 
first state close to $\pi(0)\pi(0)$ and the third state close to $\pi(\tfrac{2\pi}{L})\pi(-\tfrac{2\pi}{L})$, just as we do.  Although the 
detailed results are somewhat different, both groups obtain a second state 
below $\pi(\tfrac{2\pi}{L})\pi(-\tfrac{2\pi}{L})$, which is the crucial result.

\begin{figure}[t!]
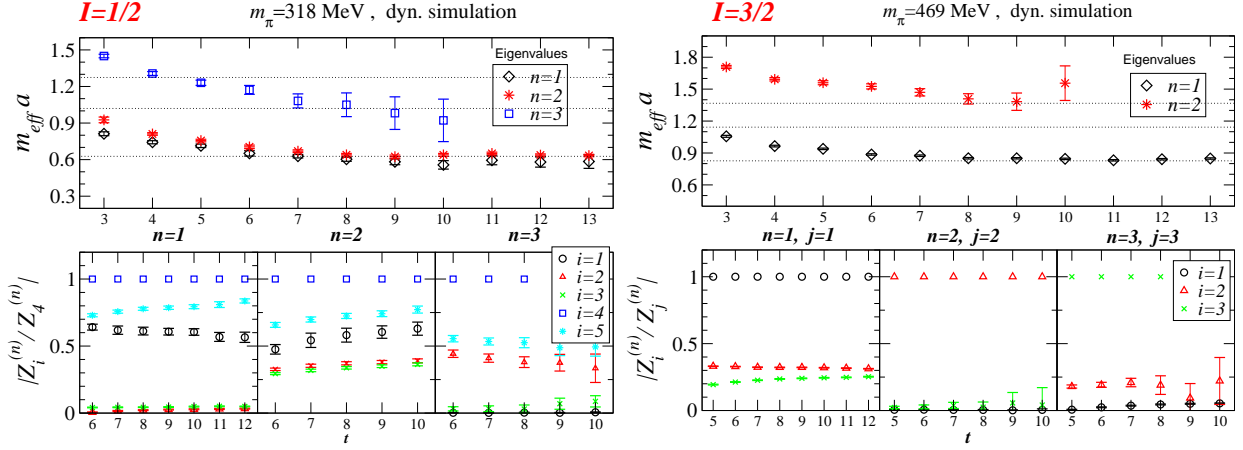

\begin{center}
\includegraphics[height=5.9cm,clip]{fig_dyn/dyn_half_C.eps} 
\includegraphics[height=5.9cm,clip]{fig_dyn/dyn_threehalf_B.eps}
\end{center}
\caption{ \small  Analogous to Fig.~\ref{fig_eff_zero_two}, but for $I=1/2$ and $I=3/2$. The  lines present the energies of non-interacting $K(k)\pi(-k)$ with $k=N {2\pi}/{L}$ and $N=0,1,\sqrt{2}$. }\label{fig_eff_half_threehalf}
\end{figure}

\subsubsection*{\bf $I=1/2$ and $I=3/2$}

The effective masses and  ratios  $|Z_i^{n}/Z_j^{n}|$ for $I=1/2,~3/2$ are shown in  Fig.~\ref{fig_eff_half_threehalf}, while the  resulting spectrum $E_n$ is given in Fig.~\ref{fig_spect_half_threehalf} and in Tables \ref{tab_results_dyn},  \ref{tab_results_q}. 

The conclusions regarding $I=1/2$ are  completely analogous to the $I=0$ case above: there is a light state in addition to $K(0)\pi(0)$ and $K(\tfrac{2\pi}{L})\pi(-\tfrac{2\pi}{L})$. Results for the exotic $I=3/2$ channel are analogous to results for  $I=2$: there is no light state in addition to $K(0)\pi(0)$ and $K(\tfrac{2\pi}{L})\pi(-\tfrac{2\pi}{L})$.

This applies to all quark masses, to both simulations and to any choice of $t_0\in [1,4]$.  We  performed the diagonalization of the $5\times 5$ and all possible $4\times 4$ and $3\times 3$ sub-matrices in $I=1/2$ case. We find that the extracted $E_{n}$ and  $|Z_i^n|$ are almost independent of these choices for all quark masses and both simulations, 
as demonstrated for a specific case in Fig.~\ref{fig_m_Z_dependence_half}.

\subsection{Volume dependence of $Z_i^n$}\label{sec_vol_dep}

For completeness we provide now the volume dependence of couplings $Z_i^n$  in the case of the quenched simulation, which was performed at two volumes $16^3\times 28$ and $12^3\times 28$ at the same lattice spacing $a=0.200(3)$ fm. We are unable to show the analogous volume dependence in the case of dynamical simulation as it was performed on a single volume. 

The expectation for the $L-$dependence of $Z_i^n(L)$ is \cite{mathur_scalar,tetra_sasa,vol_dep} 
\begin{itemize}
\item $Z_i^n(16)\simeq (\tfrac{12}{16})^{3/2}Z_i^n(12)\simeq 0.65~ Z_i^n(12)$ in case when $|n\rangle$ is two-particle state $P_1P_2$
\item $Z_i^n(16)\simeq Z_i^n(12)$ in case when $|n\rangle$ is a one-particle state (resonance) 
\end{itemize}
but these two behaviors are observed in practice only when eigenstates have very long and stable plateaus, as pointed out in \cite{alexandrou}. 

\begin{figure}[t!]
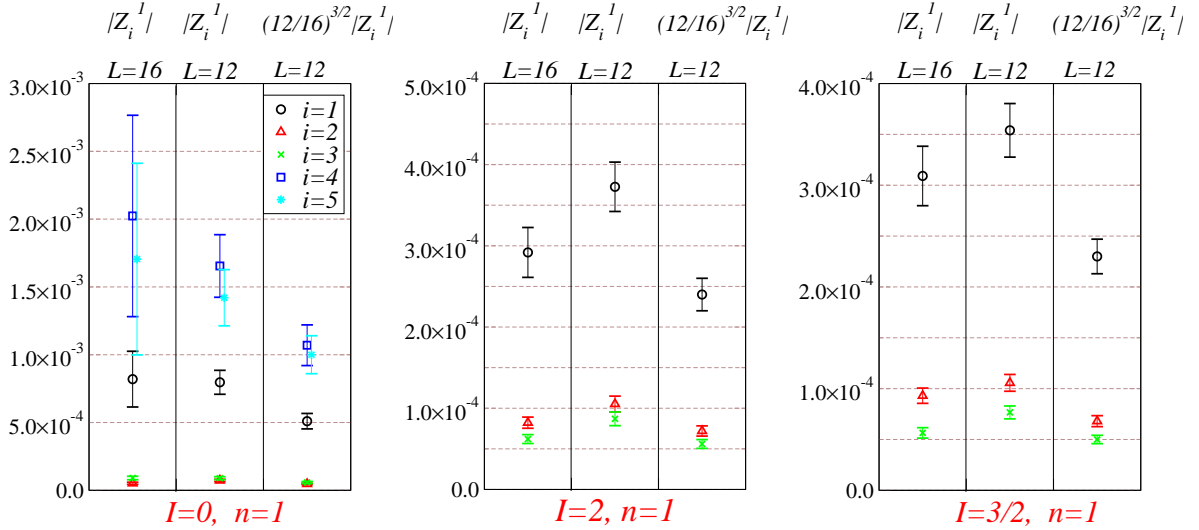

\begin{center}
\includegraphics[height=7cm,clip]{fig_q/Z_1_zero.eps}
\includegraphics[height=7cm,clip]{fig_q/Z_1_two.eps}
\includegraphics[height=7cm,clip]{fig_q/Z_1_threehalf.eps}
\end{center}
\caption{ \small
Comparison of $Z_i^n(L=16)$ with   $Z_i^n(L=12)$ and  $(\tfrac{12}{16})^{3/2}Z_i^n(L=12)$ for the ground state $n=1$,  $I=0,~2,~3/2$ and $m_\pi=342$ MeV in the quenched simulation.  The couplings for the excited states $n\geq 2$ have sizable errors and are not shown.  
All couplings  are obtained via (\ref{Z}) by a fit in the range $t\in [7,10]$  from a $3\times 3$ matrix ($I=2,~3/2$) or  a $5\times 5$ matrix ($I=0$) at $t_0=1$.  }\label{fig_vol_dep}
\end{figure} 

\begin{figure}[t!]
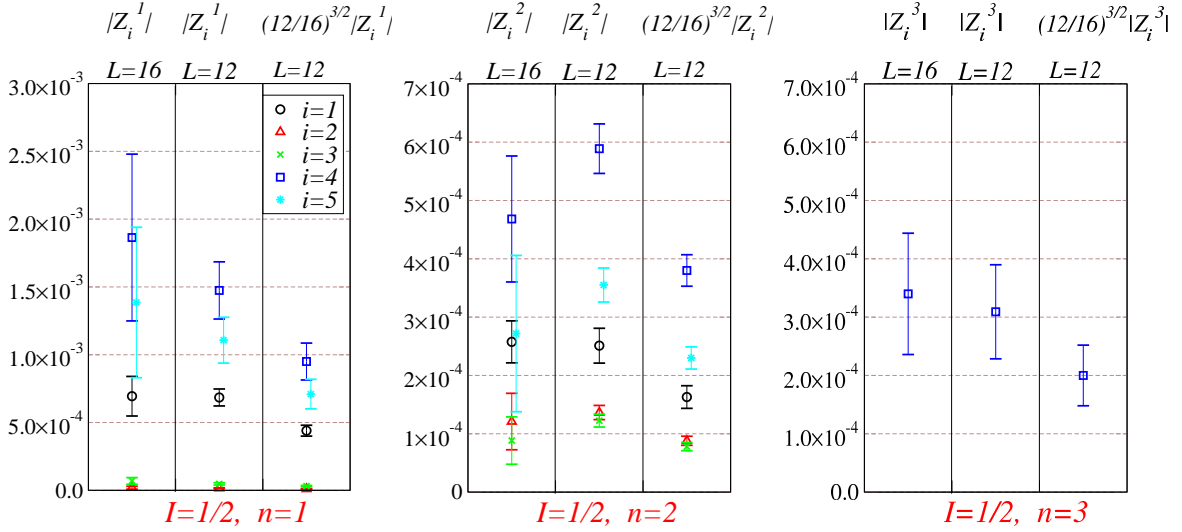

\begin{center}
\includegraphics[height=7cm,clip]{fig_q/Z_1_half.eps}
\includegraphics[height=7cm,clip]{fig_q/Z_2_half.eps}
\includegraphics[height=7cm,clip]{fig_q/Z_3_half.eps}
\end{center}
\caption{ \small   Comparison of $Z_i^n(L=16)$ with   $Z_i^n(L=12)$ and  $(\tfrac{12}{16})^{3/2}Z_i^n(L=12)$ for the three states $n=1,2,3$ and  $I=1/2$  at $m_\pi=342$ MeV in the quenched simulation. 
For the second excited state $n=3$, we show only the largest component $Z_{4}^{n=3}$, which is expected to be the most reliable. Couplings $Z$ are obtained via  (\ref{Z}) by fit in the range $t\in [7,10]$  from $5\times 5$ matrix (\ref{interpolators_zero_half}) and $t_0=1$.    }\label{fig_vol_dep_half}
\end{figure}   

Figures \ref{fig_vol_dep} and  \ref{fig_vol_dep_half}  compare 
  $Z_i^n(16)$ with $Z_i^n(12)$ and $(\tfrac{12}{16})^{3/2}Z_i^n(12)$ for all $I$. We have verified that the dependence of $Z_i^n$ on the choice of $t_0$ and interpolator set is well below the (sizable)  error-bars on $Z_i^n$ for all $i,~n,~I$. 

 The ground state $n=1$ for $I=2,3/2$ is  expected to be  $\pi\pi$ or $K\pi$ since there are no observed light resonances in these repulsive channels. We determined couplings $Z_i^{n=1}$ from the variational method (\ref{Z}) as well as from the diagonal correlators $C_{ii}(t)$  using a fit (\ref{time_scatter}) and both results agree within the error-bars.  
Our observed couplings seem to be roughly consistent with $Z_i^n(16)\simeq (\tfrac{12}{16})^{3/2}Z_i^n(12)$, which applies for two-particle states. The relation is however not satisfied exactly, which is not surprising given 
our short plateaus and sizable finite $T$ effect, as pointed out in 
\cite{alexandrou}.
 
The error-bars on extracted $Z_i^n$ for $I=0,1/2$ in Figs.~\ref{fig_vol_dep} and  \ref{fig_vol_dep_half} are to large 
to distinguish between the one- and two-particle  states\footnote{The errors on $Z_i^{n=1,2}$ for $I=0,1/2$  in 
the dynamical simulation  are smaller than in the quenched simulation (see Figs.~\ref{fig_m_Z_dependence_zero} and \ref{fig_m_Z_dependence_half}). }.

\subsection{Interpretation of the results}\label{sec_interpretation}

We now discuss our interpretation of the observed spectrum in Figs.~\ref{fig_spect_zero_two} and \ref{fig_spect_half_threehalf}, which has the 
same distinctive features in the quenched and dynamical simulations. 

\subsubsection*{\bf $I=2$ and $I=3/2$}

Let us first focus on the repulsive channels $I=2$ and $I=3/2$, where $n=1$ and 
$n=2$ states are well separated in energy. We interpret the $n=1$ ground state
 as $\pi(0)\pi(0)$ ($K(0)\pi(0)$) since it has its energy close to 
$2m_\pi$ ($m_\pi+m_K$) for $I=2$ ($I=3/2$). The time dependence of the diagonal 
correlators $C_{ii}(t)$ in Fig.~\ref{fig_c_diag} can also distinguish the 
one-particle (\ref{time_cosh}) or two-particle (\ref{time_scatter}) nature of 
the state $n=1$, that dominates the correlator\footnote{The $n=2$ state is much heavier in case of $I=2,~3/2$ and it dies out near $t\simeq T/2$.}.  The drop of cosh-type effective mass near $t\simeq T/2$ in Fig.~\ref{fig_c_diag}  speaks in favor of two-particle state $n=1$. 
The volume dependence of $Z_i^{n=1}$ couplings in Fig.~\ref{fig_vol_dep} 
is also roughly consistent with expectation for a two-particle state.    
The most important feature is a large gap between the $n=1$ and $n=2$ states, 
so we  do not observe any light resonance in $I=2,~3/2$ channel, as expected for these repulsive channels. We interpret the $n=2$ state as  $P_1(\tfrac{2\pi}{L})P_2(-\tfrac{2\pi}{L})$ scattering state. It has an  
energy somewhat above 
$E_{P_1}(\tfrac{2\pi}{L})+E_{P_2}(-\tfrac{2\pi}{L})$ (\ref{enP}), which is  most probably due to 
the small $3\times 3$ interpolator basis, which is not big enough to capture 
$E_{n=2}$ energy  well.

\subsubsection*{\bf $I=0$ and $I=1/2$}

Now we turn to the more interesting attractive channels $I=0$ and $1/2$, where 
broad resonances $\sigma$ and $\kappa$ may be expected. In each channel we 
observe two light states ($n=1,2$) near the threshold $m_{P_1}+m_{P_2}$ and 
a third state $n=3$ 
nicely consistent with  $E_{P_1}(\tfrac{2\pi}{L})+E_{P_2}(-\tfrac{2\pi}{L})$ 
(\ref{enP}). It is natural to interpret the $n=3$ state as the 
$P_1(\tfrac{2\pi}{L})P_2( -\tfrac{2\pi}{L})$ state, while the most interesting 
question is the nature of $n=1,2$ states. 
  
States $n=1$ and $n=2$ are 
orthogonal to each other according to $(\vec u^{n},C(t)\vec u^{n^\prime})\propto 
\delta_{nn^\prime}$,  so they must correspond to two distinct physical states. 
There is only one scattering state with given $|I,I_3\rangle=|0,0\rangle$ in this energy 
range,  and there is no way to envisage that the other state may also be a 
scattering state. 
This leads to a possible interpretation that the other state corresponds to a resonance $\sigma$ for $I=0$ and resonance $\kappa$ for $I=1/2$. 

\vspace{0.2cm}

Before considering this interpretation in more detail, let us point out again  
our most severe approximation\footnote{The other approximations are finite volume and lattice spacing, absence of dynamical strange quark and unphysical $u/d$ quark masses, but these approximations could not lead to the presence of unphysical scattering state near threshold.}, which amounts to omitting the disconnected 
contractions in Fig.~\ref{fig_contractions}. One might wonder if this omission could lead to the  
unphysical appearance of an additional scattering state  with wrong flavor near threshold. In the Appendix B we explicitly show that the scattering state $|I=2,I_3=0\rangle$ cannot enter as an intermediate state in our connected correlator $\langle{\cal O}^{I=0}| {\cal O}^{\dagger I=0}\rangle$, since the connected part of the matrix element $\langle 2,0|  {\cal O}^{\dagger I=0}\rangle$ vanishes.  The $|I=1,I_3=0\rangle$ state could also not appear as an intermediate state with 
$E\simeq 2m_\pi$, since there is no $\pi(0)\pi(0)$ state with $J^P=0^+$ and 
$|I=1,I_3=0\rangle$. In Appendix B we also show that the  $|I=\tfrac{3}{2},I_3=\tfrac{1}{2}\rangle$  state  cannot enter as intermediate states in our connected $I=1/2$ correlator, since the connected part of the matrix element $\langle \tfrac{3}{2},\tfrac{1}{2}|  {\cal O}^{\dagger I=1/2}\rangle$ vanishes.
Therefore we believe that the omission of the disconnected contractions cannot be responsible for the appearance of the additional light scattering state.  
Our results with two light states in the $I=0,1/2$ channels  stimulate  a future lattice simulation to search for low-lying states in $I=0,1/2$ channels using the connected as well as the disconnected 
contractions.

\vspace{0.2cm}

After these cautionary remarks, we examine the interpretation that one
 of the low-lying states is a scattering state and the other is 
$\sigma $ for $I=0$ and $\kappa$ for $I=1/2$.  According to this interpretation, the resonances $\sigma/\kappa$ found in the simulation are pure $\bar q\bar qqq$ states and have no $\bar qq$ Fock component: we are able to attribute  definite  Fock component to the simulated states  since we used tetraquark interpolators and omitted the singly and doubly disconnected contractions in Fig.~\ref{fig_contractions}. The physical resonances $\sigma/\kappa$, that correspond to the simulated states in  Nature, therefore have a $\bar q\bar qqq$ Fock component (probed in our simulation), but they may also have an additional $\bar qq$ component (not probed in our simulation without singly disconnected contractions). 

We already  mentioned that the presence of the additional light state has to be confirmed in a future simulation, which takes into account also the  disconnected contractions. Such a simulation will however not be able to claim the presence of the $\bar q\bar qqq$ Fock component due to  $\bar q\bar qqq\to \bar qq$ mixing via singly disconnected contractions.   Once the additional light state is confirmed in a simulation with disconnected contractions, the results of our present simulation will make case for the presence of $\bar q\bar qqq$ Fock components in $\sigma$ and $\kappa$.

\vspace{0.2cm}
             
Now, let us attempt to establish which one of  $n=1,2$ states is a candidate for the one-particle state $\sigma$ or $\kappa$. Based on the drop of the cosh-type effective mass for $C_{ii}(t)$ near $t\simeq T/2$ (Fig.~\ref{fig_c_diag}) we expect that a state $|n\rangle$ that dominates $C_{ii}(t)$  has a two-particle nature (\ref{time_scatter}). Our states $n=1,2$ are close to degenerate, so $C_{ii}(t)$ seems to be dominated by the state $n=1$ that has bigger coupling 
$|Z_i^{n=1}|>|Z_i^{n=2}|$ (Figs.~\ref{fig_m_Z_dependence_zero}, \ref{fig_m_Z_dependence_half},  \ref{fig_vol_dep_half}). This would indicate that $n=1$ is the scattering state $P_1(0)P_2(0)$, while $n=2$ corresponds to $\sigma$ or $\kappa$, although we cannot claim that with complete certainty. The errors on the couplings $Z_i^n(L=12,16)$ for $I=0,1/2$ are too large to distinguish  one- and two- particle behavior based on $L-$dependence of  $Z_i^n$, as noted in Section \ref{sec_vol_dep}. 

Our interpretation that $E^{n=1}$ corresponds to the scattering state $P_1(0)P_2(0)$ and $E^{n=2}=m_{\sigma,\kappa}$ corresponds to $\sigma$ and $\kappa$ is in agreement with several expectations:
\begin{itemize}
\item 
The $\sigma$ and $\kappa$ are resonances for our lower $m_\pi$ and 
so they are expected to be above the $P_1(0)P_2(0)$ scattering states, which is 
supported by the energies in Figs.~\ref{fig_spect_zero_two} and  
\ref{fig_spect_half_threehalf} at low $m_\pi$. 
\item
The $\sigma$ and $\kappa$ 
are expected to be bound states for our higher $m_\pi$ \cite{pelaez}. In the range of $m_\pi$ where this might occur,  we find $\sigma$ and $\kappa$  almost degenerate with $P_1(0)P_2(0)$, which is in qualitative agreement with expectation from the lattice study of a toy model with loosely bound and scattering states \cite{sasaki}. The authors \cite{sasaki} show that a loosely bound state lies slightly below the $m_{P_1}+m_{P_2}$ and the $P_1(0)P_2(0)$ scattering state  slightly above\footnote{The lowest scattering state $P_1(0)P_2(0)$ in the attractive channel is expected to be below $m_{P_1}+m_{P_2}$ if there is no bound state.} $m_{P_1}+m_{P_2}$, but our study is not accurate enough to reliably extract these small energy shifts. 
\item
Our resulting $m_\pi$-dependence of  $m_{\sigma,\kappa}(m_\pi)=E^{n=2}(m_\pi)$ is in qualitative agreement with the prediction of $m_{\sigma,\kappa}(m_\pi)$ within unitarized ChPT \cite{pelaez}. The authors \cite{pelaez} predict that $m_{\sigma,\kappa}$ are very close to threshold for $m_\pi/m_\pi^{phy}\in [2,3]$, which is supported by our results in Figs.~\ref{fig_spect_zero_two} and  
\ref{fig_spect_half_threehalf}. The authors \cite{pelaez} 
 find that $\sigma$ and $\kappa$ 
  transform from a resonance to a bound state at about $m_\pi\simeq 350$ MeV, 
which corresponds to intermediate $m_\pi$ in our simulation. We note that 
at $m_\pi>m_\pi^{phy}$ used in our simulation, the $\sigma$ and $\kappa$ in are expected to be significantly narrower than the broad resonances observed in the  experiment \cite{pelaez}. 
\end{itemize}

\section{Conclusions and outlook}

We determined the energy spectrum  and the couplings $\langle 0|{\cal O}_i|n\rangle$  for the states with $J^{PC}=0^{++}$, isospin $I=0,~1/2,~3/2,~2$ and $\vec p=\vec 0$  using  a number of tetraquark interpolators $\bar q\bar q qq$ at the source and the sink. We omitted the disconnected contractions in Fig.~\ref{fig_contractions}.           
Our main question is whether there are any light states in addition to the towers of scattering states $P_1(k)P_2(-k)$ with $k=0,2\pi /L,..$ and $P_1P_2=\pi\pi$ or $K\pi$. Such additional states could be related to resonances $\sigma$ or $\kappa$ with a sizable tetraquark component. 

The resulting spectra in Figs.~\ref{fig_spect_zero_two} and \ref{fig_spect_half_threehalf} show qualitative agreement between the dynamical and the quenched simulations.  In the repulsive channels $I=2,~3/2$, where no resonance is expected, we indeed find only the scattering states $P_1(0)P_2(0)$ and $P_1(\tfrac{2\pi}{L}) P_2(-\tfrac{2\pi}{L})$ with no additional light state. In the attractive channels  $I=0,~1/2$ we find two (orthogonal) states close to the threshold $m_{P_1}+m_{P_2}$ and another state consistent with $P_1(\tfrac{2\pi}{L}) P_2(-\tfrac{2\pi}{L})$, so we do find an additional light state.  
This leads to a possible interpretation that one of the two light states is a scattering state $P_1(0)P_2(0)$ and the other one corresponds to a resonance $\sigma$ for $I=0$ and resonance $\kappa$ for $I=1/2$. According to this interpretation, the physical  resonances $\sigma$ and $\kappa$   have a nonzero $\bar q\bar qqq$ Fock component, since the corresponding states  in our simulation couple to the tetraquark interpolators.  Along these lines, these physical resonances could not be pure $\bar q q$ resonances since pure $\bar q q$  resonances do not couple to tetraquark interpolators in absence of singly  disconnected diagrams in our simulation. Our candidates for $\sigma$ and $\kappa$ have 
an $m_\pi$ dependence in qualitative agreement with expectation from unitarized ChPT \cite{pelaez}.

The volume dependence of the couplings $\langle 0|{\cal O}_i|n\rangle$ for $I=2,3/2$ is roughly consistent with the above interpretation, while errors on $I=0,1/2$ couplings are too large to distinguish the one- and two-particle states based on this criterion. 
 We also use the time-dependence of the correlators and eigenvalues at finite temporal extent as criteria for distinguishing the one or two particle states, and we demonstrate that one of the two light states in the $I=0,~1/2$ channels is a scattering state. Along the way, we derive the analytical conditions, which have to be satisfied so that the eigenvalues corresponding to  the one-particle states   would have a simple time dependence  proportional to $e^{-E_nt}+e^{-E_n(T-t)}$. 

We explored the possibility whether the omission of the disconnected contractions could lead to ``unphysical'' light eigenstate with $I=2$ in the $I=0$ channel. 
We explicitly verified that $I=2$ state cannot enter as intermediate state 
in  our connected $I=0$ correlator.  Similarly, $I=3/2$ state cannot enter as intermediate state 
in  our connected $I=1/2$ correlator.

The ultimate method to study $\sigma$ and $\kappa$ on the lattice would involve the study of the spectrum and couplings in presence of the disconnected contractions  and the  $\bar q\bar qqq\leftrightarrow \bar qq\leftrightarrow vac\leftrightarrow glue$ mixing, using interpolators that cover these  Fock components. The recently proposed distillation method \cite{distillation} could prove useful for determining the correlators with $P_1(\vec k)P_2(-\vec k)$ interpolators or disconnected contractions.
Such a study has to be done as a function of lattice size $L$ in order to extract the resonance  mass and width  using the L\"{u}scher's finite volume method \cite{luscher,vol_dep, sasaki}.


\vspace{1cm}
    
{\bf Acknowledgments}

\vspace{0.2cm}
 
We would like to thank W. Detmold, R. Edwards, G. Engel,  C. Gattringer,  J. Juge,  M. Komelj, C. Morningstar, J. Pelaez, S. Sasaki and  M. Savage  for valuable discussions. The configurations with dynamical quarks have been produced by the
BGR-collaboration  on the SGI Altix 4700 of the Leibniz-Rechenzentrum Munich. The quenched part of simulation was done at NERSC, USA. This work is supported by the Slovenian Research Agency, by the European RTN network FLAVIAnet (contract number MRTN-CT-035482), by the Slovenian-Austrian bilateral project (contract number  BI-AT/09-10-012), the USA DOE Grant DE-FG05-84ER40154, the Austrian grant FWF DK W1203-N08, the German DFG grant SFB-TR55 and by Natural Sciences and Engineering Research Council of Canada.


\newpage
\appendix

\section*{ Appendix A: Effect of finite $T$ on eigenvalues of the generalized eigenvalue method}

\vspace{0.3cm}

Here we consider   the effect of finite $T$ on the eigenvalues of the generalized eigenvalue problem. We take a simple example with only two physical states of  given quantum numbers $J^{PC}$ and $I$: a one-particle state $A$ (for example $\sigma$ or $\kappa$) and a two-particle state $P_1P_2$ (for example $\pi\pi$ or $K\pi$). Let us study two eigenvalues of the $2\times 2$ correlation matrix (for example using two interpolators from (\ref{interpolators_zero_half}))
 \begin{equation}
C_{ij}(t)=\frac{1}{Z_T}Tr[e^{-H T} {\cal O}_i(t){\cal O}_j^\dagger (0)]=\frac{1}{Z_T}\sum_{m,n} \langle m|e^{-H(T-t)}{\cal O}_i|n\rangle\langle n|e^{-Ht}{\cal O}_j^\dagger|m\rangle 
\end{equation}
and the relevant states $n,m=A$, $P_1P_2$, $P_1$, $P_2$ giving \cite{tetra_sasa}
\begin{align}
\label{cor_app}
C_{ij}(t)=&\ \langle 0|{\cal O}_i |A\rangle\langle A | {\cal O}_j^\dagger |0\rangle e^{-m_A t}+\langle A^\dagger |{\cal O}_i |0\rangle\langle 0 | {\cal O}_j^\dagger |A^\dagger\rangle e^{-m_A (T-t)}\\
  +&\ \langle 0|{\cal O}_i |P_1P_2\rangle\langle P_1P_2 | {\cal O}_j^\dagger |0\rangle e^{-E_{P_1P_2} t}+\langle P_1^\dagger P_2^\dagger |{\cal O}_i |0\rangle\langle 0 | {\cal O}_j^\dagger |P_1^\dagger P_2^\dagger \rangle e^{-E_{P_1P_2} (T-t)}\nonumber\\
+&\ \langle P_1^\dagger|{\cal O}_i |P_2\rangle\langle P_2 | {\cal O}_j^\dagger |P_1^\dagger \rangle e^{-E_{P_1} (T-t)}e^{-E_{P_2} t}+\langle P_2^\dagger|{\cal O}_i |P_1 \rangle\langle P_1 | {\cal O}_j^\dagger |P_2^\dagger \rangle e^{-E_{P_2} (T-t)}e^{-E_{P_1} t}\nonumber\\
=&\ \langle 0|{\cal O}_i |A\rangle\langle 0 | {\cal O}_j |A\rangle^* ~\bigl[ e^{-m_A t}\pm e^{-m_A (T-t)}\bigr]\nonumber \\
  +  &\ \langle 0|{\cal O}_i |P_1P_2\rangle\langle 0 | {\cal O}_j  |P_1P_2\rangle^*~ \bigl[ e^{-E_{P_1P_2} t}\pm e^{-E_{P_1P_2} (T-t)}\bigr]\nonumber \\
+&\ \langle P_1^\dagger|{\cal O}_i |P_2\rangle\langle P_1^\dagger | {\cal O}_j |P_2 \rangle^*~\bigl[ e^{-E_{P_1} (T-t)}e^{-E_{P_2} t}\pm e^{-E_{P_2} (T-t)}e^{-E_{P_1} t}\bigr]\nonumber~.
\end{align}
The signs ``$\pm$'' depend on the symmetry properties of the interpolators and states. We will assume ``$+$'' sign everywhere, since all our diagonal and non-diagonal correlators in the actual simulation are symmetric with respect to $t\leftrightarrow T-t$ (if the ``$-$'' sign occurs in some terms, the ``backward propagation'' described in Section IIf of \cite{ci_derivative} may occur). 

If the coefficients in the third line were proportional to the coefficients in the second line 
\begin{equation}
\label{condition_app}
\langle P_1^\dagger|{\cal O}_i|P_2\rangle= R_i ~\langle 0|{\cal O}_i|P_1P_2\rangle\qquad {\mathrm {with}\ R_i=R~,\ {\mathrm {independent\ of}}\ i}~, 
\end{equation}
the third line would depend linearly on the second line 
\begin{equation}
C_{ij}(t)=\sum_{n=1,2}Z_i^nZ_j^{n*} f^n(t)
 \end{equation}
with 
\begin{align}
Z_i^1&=\langle 0|{\cal O}_i|A\rangle,    \   \qquad f^1(t)=e^{-m_A t}+ e^{-m_A (T-t)}~,\nonumber\\
Z_i^2&=\langle 0|{\cal O}_i|P_1P_2\rangle, \quad f^2(t)=e^{-E_{P_1P_2} t}+ e^{-E_{P_1P_2} (T-t)}+R\bigl[e^{-E_{P_1} (T-t)}e^{-E_{P_2} t}+ e^{-E_{P_2} (T-t)}e^{-E_{P_1} t}\bigr].
\end{align}
 The eigenvalues of  $C(t)\vec u^{n}(t)=\lambda^{n}(t,t_0) C(t_0)\vec u^{n}(t)$ can be obtained in this case following the same steps as  in \cite{ci_ghosts}
\begin{equation}
\lambda^n(t)=\frac{f^n(t)}{f^n(t_0)}~. 
\end{equation}
So, only if the condition (\ref{condition_app}) would be exactly satisfied, the following applies:  
\begin{enumerate}
\item[(i)] the eigenvalue corresponding to the one-particle state  would have a time-dependence proportional to $f^1(t)$  and
\item[(ii)] the  eigenvalue corresponding to the two-particle state would have a time-dependence proportional to $f^2(t)$. 
\end{enumerate}
However, we believe that  condition (\ref{condition_app}) is {\it not} exactly satisfied, as argued after (\ref{time_scatter}) in the main text. We have not  rigorously derived the eigenvalues for this case. Based on the numerical example illustrated below, we however expect that eigenvalues have the form 
\begin{equation}
\label{time_lam_app}
\lambda^{n}(t)= w^n~[e^{-E_nt}+e^{-E_n(T-t)}]+\tilde w^n ~[e^{-m_{P_1}t}e^{-m_{P_2}(T-t)}+ e^{-m_{P_2}t}e^{-m_{P_1}(T-t)}]~ 
\end{equation}
for two-particle states as well as for one-particle states. 
  
\begin{figure}[t!]
\begin{center}
\includegraphics[height=5cm,clip]{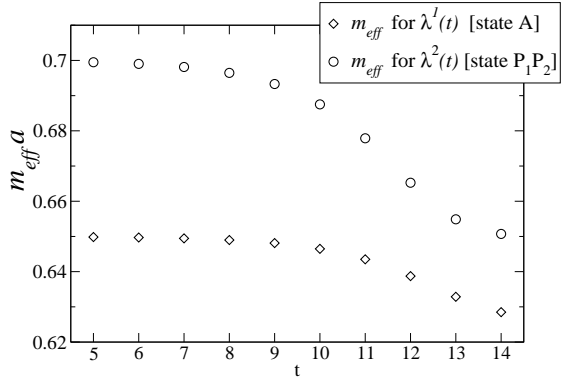}
\end{center}
\caption{ \small Cosh-type effective masses (\ref{eff_cosh}) for  two eigenvalues of the generalized eigenvalue problem with $2\times 2$ correlation matrix (\ref{cor_app},\ref{condition_app}) and $R_1\not=R_2$. A specific choice of the parameters is given in the text. }\label{fig_toy_model}
\end{figure}

Let us demonstrate that on a specific example with $m_A=0.6$, $m_{P_1}=0.3$, $m_{P_2}=0.4$, $m_{P_1P_2}=0.7$, $Z_{1}^1=Z_{2}^2=\cos(0.2)$, $Z_1^2=-Z_2^1=\sin(0.2)$, $T=30$, $R_1=2$, $R_2=0.4$ and $t_0=4$. We find the two eigenvalues for this case. Fig.~\ref{fig_toy_model}  shows the corresponding cosh-type effective masses, defined in (\ref{eff_cosh}). The effective masses for the two-particle state $P_1P_2$ as well as for the one particle state $A$ drop near $t\simeq T/2$. So none of them has the time-dependence  $e^{-Et}+e^{-E(T-t)}$, but a more complicated time dependence (\ref{time_lam_app}).   The eigenvalue corresponding to the state $A$ is proportional to $e^{-m_At}+e^{-m_A(T-t)}$ only in the special case $R_1=R_2$. 
To demonstrate that the time-dependence of $\lambda^n(t)$ is of the form (\ref{time_lam_app}) to a very good approximation, we fit the resulting $\lambda^n(t)$ to (\ref{time_lam_app}) with three free parameters $E_n,w^n,~\tilde w^n$. The resulting $E_n$ agree with input $m_A$ or $E_{P_1P_2}$ almost exactly and the resulting $\chi^2$ of the fit\footnote{We attribute artificial error-bars $\sigma^n(t)=\lambda^n(t)/10$ to the values of $\lambda^n(t)$, which enter uncorrelated $\chi^2$.} is extremely small. 

\section*{ Appendix B: Discussion concerning omission of disconnected diagrams in $I=0,1/2$ correlators}

In this appendix we show that  the scattering state $|I=2,I_3=0\rangle$  cannot enter as an intermediate state in the connected correlation function $\langle {\cal O}^{I=0}|{\cal O}^{\dagger I=0}\rangle$, while the scattering  state $|I=\tfrac{3}{2},I_3=\tfrac{1}{2}\rangle$ cannot enter in  $\langle {\cal O}^{I=1/2}|{\cal O}^{\dagger I=1/2}\rangle$.

\subsection*{{\it\bf I=0}}

The interpolators ${\cal O}^{I=0}_{1,..,3}$ (\ref{interpolators_zero_half}) 
\begin{equation*}
{\cal O}_{1,..,3}^{I=0}=-\tfrac{1}{\sqrt{3}}\bigl[2(\bar d\Gamma u)(\bar u\Gamma d)+\tfrac{1}{2}(\bar u\Gamma u)(\bar u\Gamma u)+\tfrac{1}{2}(\bar d\Gamma d)(\bar d\Gamma d)-(\bar u\Gamma u)(\bar d\Gamma d)\bigr]\nonumber
\end{equation*}
are  $|I=0,I_3=0\rangle$ Clebsh-Gordon combinations of two $I=1$ fields: 
$|0,0\rangle=\tfrac{1}{\sqrt{3}}\{
2|1,1\rangle |1, -1\rangle -|1, 0\rangle |1,0\rangle  \}=\tfrac{1}{\sqrt{3}}\{-2(\bar d\Gamma u)(\bar u\Gamma d)-\tfrac{1}{2}[(\bar d\Gamma d)-(\bar u\Gamma u)]^2\}$, where we used fields $\bar q\Gamma q'$ with definite $|I,I_3\rangle$ from Table \ref{tab_CG}. The $|I=2,I_3=0\rangle$ state has flavor structure 
\begin{equation*}
|2,0\rangle=\tfrac{1}{\sqrt{6}}\{
2|1,1\rangle |1, -1\rangle +2|1, 0\rangle |1,0\rangle\}
=\tfrac{1}{\sqrt{6}}\bigl\{-2(\bar d\Gamma' u)(\bar u\Gamma' d)+[(\bar d\Gamma' d)-(\bar u\Gamma' u)]^2\}~.
\end{equation*} 
We need to find out whether the state $| 2,0\rangle$ can enter as an intermediate state in the connected part of the $I=0$ correlation function, therefore we evaluate the connected part of the matrix element 
\begin{align}
\label{appB}
\langle 2,0|{\cal O}_{1,..,3}^{\dagger I=0}\rangle_{con}&=\tfrac{1}{\sqrt{18}}\ \bigl\langle 2(\bar d\Gamma' u)(\bar u\Gamma' d)-(\bar u\Gamma' u)(\bar u\Gamma' u)-(\bar d\Gamma' d)(\bar d\Gamma' d)+2(\bar u\Gamma' u)(\bar d\Gamma' d) |\\
&\qquad\qquad  2(\bar d\Gamma u)(\bar u\Gamma d)+\tfrac{1}{2}(\bar u\Gamma u)(\bar u\Gamma u)+\tfrac{1}{2}(\bar d\Gamma d)(\bar d\Gamma d)-(\bar u\Gamma u)(\bar d\Gamma d)\bigr\rangle \nonumber\\
&=\tfrac{1}{\sqrt{18}}\ \{4D(t)+[C(t)-D(t)]+[C(t)-D(t)]-2D(t)+2C(t)-4C(t)\}=0~.\nonumber
\end{align} 
Here $D(t)$ and $C(t)$ denote the ``direct'' and ``crossed'' connected contractions in the notation of \cite{tetra_jaffe}, while six terms in the third line of (\ref{appB}) refer to  non-zero contractions of separate terms $aa^\prime,bb^\prime,cc^\prime,dd^\prime,ad^\prime,da^\prime$; here $abcd$ refer to terms in the first row (\ref{appB}) and $a^\prime b^\prime c^\prime d^\prime$ refer to the second row. So we find that $| 2,0\rangle$ state cannot couple to ${\cal O}_{1,..,3}^{I=0}$ via connected contractions. 

It is straightforward to show  that  $\langle 2,0|{\cal O}_{4,5}^{\dagger I=0}\rangle_{con}=0$ also for the  diquark anti-diquark interpolators ${\cal O}_{4,5}^{I=0}$  (\ref{interpolators_zero_half}). 
This is due to the cancellation between results from two terms in $|2,0\rangle \propto (\bar d\Gamma' u)(\bar u\Gamma' d)+(\bar d\Gamma' d)(\bar u\Gamma' u)+..~$, which 
have the flavor structure of ${\cal O}_{4,5}^{I=0}\simeq \bar u \bar d ud$.

So, the connected part of $\langle 2,0|{\cal O}^{\dagger I=0}\rangle$ vanishes for all our interpolators and the  $|I=2,I_3=0\rangle$ state cannot enter as an intermediate state in our connected   $I=0$ correlator.

\subsection*{{\it\bf I=1/2}}

Now let's see whether state $|I=\tfrac{3}{2},I_3=\tfrac{1}{2}\rangle$
\begin{equation*}
|\tfrac{3}{2},\tfrac{1}{2}\rangle=\tfrac{1}{\sqrt{3}}\{
 |\tfrac{1}{2}, -\tfrac{1}{2}\rangle|1,1\rangle +\sqrt{2} |\tfrac{1}{2},\tfrac{1}{2}\rangle|1, 0\rangle\}
=\tfrac{1}{\sqrt{3}}\{
(\bar s\Gamma' d)(\bar d\Gamma' u)+(\bar s\Gamma' u)[(\bar d\Gamma' d)-(\bar u\Gamma' u)]\}
\end{equation*} 
can enter as an intermediate state of connected $I=1/2$ correlator with ${\cal O}_{1,..,3}^{I=1/2}$ (\ref{interpolators_zero_half})
\begin{equation*}
{\cal O}_{1,..,3}^{I=1/2}=\sum_{q=u,d,s}(\bar s\Gamma q)(\bar q\Gamma u)= (\bar s\Gamma u)(\bar u\Gamma u)+(\bar s\Gamma d)(\bar d\Gamma u)+(\bar s\Gamma s)(\bar s\Gamma u)~.
\end{equation*}
The corresponding connected part of the matrix element is
\begin{align*}
\langle \tfrac{3}{2},\tfrac{1}{2}|{\cal O}_{1,..,3}^{\dagger I=1/2}\rangle_{con}&=\tfrac{1}{\sqrt{3}}\ \bigr\langle  (\bar s\Gamma' d)(\bar d\Gamma' u)-(\bar s\Gamma' u)(\bar u\Gamma' u)+(\bar s\Gamma' u)(\bar d\Gamma' d) |\nonumber \\
&\qquad\qquad (\bar d\Gamma s)(\bar u\Gamma d)+(\bar u\Gamma s)(\bar u\Gamma u)+(\bar s\Gamma s)(\bar u\Gamma s)\bigr\rangle\nonumber\\
&=\tfrac{1}{\sqrt{3}}~\{D(t)+[C(t)-D(t)]-C(t)\}=0 
\end{align*}
where three  terms  refer to  non-zero contractions of separate terms $aa^\prime,bb^\prime,ca^\prime$. 
So $| \tfrac{3}{2},\tfrac{1}{2}\rangle$ state cannot couple to ${\cal O}_{1,..,3}^{I=1/2}$ via connected contractions.

We also find that  $\langle\tfrac{3}{2},\tfrac{1}{2} |{\cal O}_{4,5}^{\dagger I=1/2}\rangle_{con}=0$ for diquark anti-diquark interpolators ${\cal O}_{4,5}^{I=1/2}$   (\ref{interpolators_zero_half}). 
This is due to the cancellation between results from two terms in $\langle\tfrac{3}{2},\tfrac{1}{2}|\propto (\bar s\Gamma' d)(\bar d\Gamma' u)+(\bar s\Gamma' u)(\bar d\Gamma' d)+.. ~$, which 
have the flavor structure of ${\cal O}_{4,5}^{I=1/2}\simeq \bar s \bar d du$.

So, 
the connected part of $\langle \tfrac{3}{2},\tfrac{1}{2}|{\cal O}^{\dagger I=1/2}\rangle$ vanishes for all our interpolators and the  $|I=\tfrac{3}{2},I_3=\tfrac{1}{2}\rangle$ state cannot enter as an intermediate state in our connected   $I=1/2$ correlator.

\vspace{2cm}


\begin{table}[h]
\begin{center}
\begin{tabular}{c c c c c c c c}
ensemble & $a$~[fm] & $m_{\pi}~$[MeV] & $m_\pi a$ & $m_K a$ & $N$ &  $\kappa$ & $conf.$ \\ 
\hline
C & $0.1440(12)$ & $318(5)$ & $0.232(4)$ & $0.391(3)$ & $15$ & $0.223$ & 200\\
B & $0.1500(12)$ & $469(4)$ & $0.357(3)$ & $0.462(3)$ & $15$ & $0.222$ & 200 \\
A & $0.1507(17)$ & $526(7)$ & $0.402(5)$ & $0.465(3)$ & $17$ & $0.212$ & 100\\
\hline
\end{tabular}
\end{center}
\caption{ \small Two-flavor dynamical  ensembles with Chirally Improved quarks \cite{ci_dyn}:  pseudoscalar masses, Jacobi-smearing parameters ($N$, $\kappa$) and the number of configurations are listed. All ensembles have volume $16^3\times 32$. }\label{tab_run_dyn}
\end{table}

\begin{table}[h]
\begin{center}
\begin{tabular}{c c c c }
$m_{\pi}~$[MeV] & $m_\pi a$ &  $m_K a$ & $conf.$ \\ 
\hline
$230(7)$ & $0.2332(56)$ & $0.515(3)$ & 300\\
$342(6)$ & $0.3470(40)$ & $0.545(2)$ & 300\\
$478(8)$ & $0.4840(32)$ & $0.596(2)$ & 300\\
\hline
\end{tabular}
\end{center}
\caption{ \small The quenched simulation with overlap quarks is performed at two volumes $16^3\times 28$ and $12^3\times 28$ at the same lattice spacing $a=0.200(3)$ fm \cite{chiral_logs}. The analysis is done at three pion masses above (these $m_\pi$ are determined at the larger volume). }\label{tab_run_q}
\end{table}


\begin{table}[h]
\begin{center}
\begin{tabular}{c c| c c c c c c c c } 
 $I$ &  $n$ & $m_\pi~$[MeV]  & $E_na$ & $w^n$ & $\tilde w^n$ & $t_0$ & interp. & {\small $t_{min}-t_{max}$} & $\chi^2$/dof\\   
\hline
\hline
$0$ & $1$ & $318$ & $0.36(4)$ & $0.33(8)$ &$-0.15(63)$ & $1$ & ${\cal O}_{12345}$ & $8-15$ & $0.015$ \\
$0$ & $1$ & $469$ & $0.72(1)$ & $6.6(6)$  & $8.7(1.0)$  & $3$ & ${\cal O}_{12345}$ & $8-15$ & $0.0065$ \\
$0$ & $1$ & $526$ & $0.81(1)$ & $8.7(7)$  & $9.2(9)$   & $3$ & ${\cal O}_{12345}$ & $8-15$ & $0.14$ \\
\hline
$0$ & $2$ & $318$ & $0.41(2)$ &$0.061(10)$& $0.049(16)$& $1$ & ${\cal O}_{12345}$ & $8-15$ & $0.088$ \\
$0$ & $2$ & $469$ & $0.72(1)$ & $3.8(3)$  & $5.3(4)$   & $3$ & ${\cal O}_{12345}$ & $8-15$ & $0.024$ \\
$0$ & $2$ & $526$ & $0.82(1)$ & $5.3(5)$  & $6.6(5)$   & $3$ & ${\cal O}_{12345}$ & $8-15$ & $0.018$ \\
\hline
$0$ & $3$ & $318$ & $0.93(9)$ & $0.23(12)$&            & $1$ & ${\cal O}_{12345}$ & $6-9$ & $0.36$ \\
$0$ & $3$ & $469$ & $1.11(5)$ & $10.5(3.7)$&           & $3$ & ${\cal O}_{12345}$ & $7-9$ & $0.28$ \\
$0$ & $3$ & $526$ & $1.07(6)$ & $7.7(2.9)$ &           & $3$ & ${\cal O}_{12345}$ & $6-10$ & $1.1$ \\
\hline
\hline
$2$ & $1$ & $318$ & $0.45(2)$& $0.33(4)$  & $0.32(8)$ & $1$ & ${\cal O}_{123}$   & $8-15$ & $0.0086$ \\
$2$ & $1$ & $469$ & $0.736(5)$& $0.47(2)$  & $0.54(3)$ & $1$ & ${\cal O}_{123}$   & $8-15$ & $0.016$ \\
$2$ & $1$ & $526$ & $0.827(7)$& $0.57(4)$  & $0.54(4)$ & $1$ & ${\cal O}_{123}$   & $8-15$ & $0.0084$ \\
\hline 
$2$ & $2$ & $318$ & $1.33(19)$& $0.73(91)$ &            & $1$ & ${\cal O}_{123}$   & $6-9$  & $0.0039$\\ 
$2$ & $2$ & $469$ & $1.33(4)$ & $0.41(13)$ &           & $1$ & ${\cal O}_{123}$   & $7-10$ & $0.25$ \\
$2$ & $2$ & $526$ & $1.35(8)$ & $0.40(27)$ &           & $1$ & ${\cal O}_{123}$   & $8-11$ & $0.60$ \\
\hline
\hline
$1/2$&$1$ & $318$ & $0.55(3)$ & $2.8(6)$   & $1.5(1.4)$& $3$ & ${\cal O}_{12345}$ & $8-15$ & $0.077$ \\
$1/2$&$1$ & $469$ & $0.826(8)$& $8.1(6)$   & $9.2(9)$  & $3$ & ${\cal O}_{12345}$ & $8-15$ & $0.032$ \\
$1/2$&$1$ & $526$ & $0.87(1)$ & $9.4(9)$   & $9.4(9)$  & $3$ & ${\cal O}_{12345}$ & $8-15$ & $0.11$ \\
\hline
$1/2$&$2$ & $318$ & $0.636(8)$& $3.4(2)$   & $3.2(3)$  & $3$ & ${\cal O}_{12345}$ & $8-15$ & $0.011$ \\ 
$1/2$&$2$ & $469$ & $0.852(5)$& $6.7(3)$   & $7.2(4)$  & $3$ & ${\cal O}_{12345}$ & $8-15$ & $0.048$ \\
$1/2$&$2$ & $526$ & $0.907(7)$& $8.3(6)$   & $7.4(4)$  & $3$ & ${\cal O}_{12345}$ & $8-15$ & $0.012$ \\
\hline
$1/2$&$3$ & $318$ & $1.05(7)$ & $8.8(40)$  &           & $3$ & ${\cal O}_{12345}$ & $7-10$ & $0.11$ \\
$1/2$&$3$ & $469$ & $1.16(6)$ & $9.9(45)$  &           & $3$ & ${\cal O}_{12345}$ & $8-10$ & $0.57$ \\
$1/2$&$3$ & $526$ & $1.14(4)$ & $7.9(23)$  &           & $3$ & ${\cal O}_{12345}$ & $7-10$ & $0.32$ \\
\hline
\hline
$3/2$&$1$ & $318$ & $0.630(9)$ & $0.36(2)$  &  $0.34(4)$ & $1$ & ${\cal O}_{123}$   & $8-15$ & $0.0088$ \\
$3/2$&$1$ & $469$ & $0.844(5)$ & $0.50(2)$  &  $0.55(3)$ & $1$ & ${\cal O}_{123}$   & $8-15$ & $0.042$ \\
$3/2$&$1$ & $526$ & $0.896(6)$ & $0.60(4)$  &  $0.57(4)$ & $1$ & ${\cal O}_{123}$   & $8-15$ & $0.013$ \\
\hline
$3/2$&$2$ & $318$ & $1.27(9)$ & $0.33(21)$ &          & $1$ & ${\cal O}_{123}$   & $7-10$ & $0.62$ \\
$3/2$&$2$ & $469$ & $1.40(5)$  & $0.50(19)$ &          & $1$ & ${\cal O}_{123}$   & $8-10$ & $0.017$ \\ 
$3/2$&$2$ & $526$ & $1.34(9)$  & $2.5(2)$   &          & $1$ & ${\cal O}_{123}$   & $9-11$ & $0.058$ \\    
\hline
\hline
\end{tabular}
\end{center}
\caption{ \small   Extracted energies $E_na$ ($a^{-1}\simeq 1.3~$GeV) together with $w^n$  in the dynamical simulation for all isospins. The fit form (\ref{time_lam}) is used whenever $\tilde w_n$ is provided, while fit form (\ref{time_lam_naive}) is used where $\tilde w^n$ is not provided. The interpolator basis, $t_0$, fit ranges and uncorrelated $\chi^2/$(degrees\ of\ freedom)
 are also presented. 
The  $n=1$ states  with $I=0,~1/2$ and the lowest $m_\pi$ have badly determined $\tilde w^n$ since they have almost flat cosh-type effective mass, which indicates they are roughly consistent with  (\ref{time_lam_naive}) and $\tilde w\simeq 0$ (finite $T$ effect is less significant  at low $m_\pi$ \cite{tetra_sasa}).     }\label{tab_results_dyn}
\end{table}

\begin{table}[h]
\begin{center}
\begin{tabular}{c c| c c c c c c c c }
%
 $I$ &  $n$ & $m_\pi~$[MeV]  & $E_na$ & $w^n$ & $\tilde w^n$ & $t_0$ & interp. & {\small $t_{min}-t_{max}$} & $\chi^2$/dof\\   
\hline
\hline
$0$   & $1$ & $230$ &  $0.40(3)$ & $1.2(6)$  &          & $3$ &  ${\cal O}_{123}$   & $8-10$ & $0.0013$\\
$0$   & $1$ & $342$ &  $0.72(2)$ & $3.1(6)$  & $1.5(4)$ & $3$ &  ${\cal O}_{12345}$ & $8-13$ & $0.0010$\\
$0$   & $1$ & $478$ &  $1.03(2)$ & $7.2(14)$ & $2.0(3)$ & $3$ &  ${\cal O}_{1245}$  & $8-13$ & $0.0020$\\
\hline
$0$   & $2$ & $230$ &  $0.45(4)$ & $0.15(6)$ &          & $3$ &  ${\cal O}_{123}$   & $8-10$ & $0.039$ \\
$0$   & $2$ & $342$ &  $0.75(8)$ & $0.45(44)$& $0.57(9)$& $3$ &  ${\cal O}_{12345}$ & $8-13$ & $0.0084$\\
$0$   & $2$ & $478$ &  $1.03(6)$ & $0.46(43)$& $0.38(4)$& $3$ &  ${\cal O}_{1245}$  & $8-13$ & $0.018$ \\
\hline
$0$   & $3$ & $230$ &  $1.05(21)$& $0.8(12)$ &          & $3$ &  ${\cal O}_{123}$   & $6-8$  & $0.41$   \\
$0$   & $3$ & $342$ &  $1.15(6)$ & $1.3(5)$  &          & $3$ &  ${\cal O}_{12345}$ & $7-9$  & $0.018$  \\
$0$   & $3$ & $478$ &  $1.33(5)$ & $1.2(5)$  &          & $3$ &  ${\cal O}_{1245}$  & $8-10$ & $0.053$  \\
\hline
\hline
$2$   & $1$ & $230$ &  $0.54(9)$ & $3.3(19)$ & $2.5(6)$ & $3$ &  ${\cal O}_{123}$   & $8-13$ & $0.085$  \\
$2$   & $1$ & $342$ &  $0.719(9)$& $1.8(2)$  & $1.64(9)$& $3$ &  ${\cal O}_{123}$   & $8-13$ & $0.0027$ \\
$2$   & $1$ & $478$ &  $1.032(7)$& $2.8(2)$  & $1.90(7)$& $3$ &  ${\cal O}_{123}$   & $8-13$ & $0.16$  \\
\hline
$2$   & $2$ & $342$ &  $1.00(37)$& $0.05(55)$&          & $3$ &  ${\cal O}_{123}$   & $7-10$ & $0.21$  \\
$2$   & $2$ & $478$ &  $1.50(9)$ & $0.62(52)$&          & $3$ &  ${\cal O}_{123}$   & $8-11$ & $0.21$  \\
\hline
\hline
$1/2$ & $1$ & $230$ &  $0.76(3)$ & $2.7(10)$ & $1.6(5)$ & $3$ &  ${\cal O}_{12345}$ & $9-13$ & $0.00034$\\
$1/2$ & $1$ & $342$ &  $0.94(2)$ & $4.5(8)$  & $1.7(3)$ & $3$ &  ${\cal O}_{12345}$ & $9-13$ & $0.0030$\\
$1/2$ & $1$ & $478$ &  $1.16(2)$ & $8.1(13)$ & $1.9(2)$ & $3$ &  ${\cal O}_{12345}$ & $9-13$ & $0.0038$\\
\hline
$1/2$ & $2$ & $230$ &  $0.76(3)$ & $1.1(3)$  & $1.1(2)$ & $3$ &  ${\cal O}_{12345}$ & $9-13$ & $0.0030$\\
$1/2$ & $2$ & $342$ &  $0.94(2)$ & $1.2(2)$  & $0.90(6)$& $3$ &  ${\cal O}_{12345}$ & $9-13$ & $0.00018$\\
$1/2$ & $2$ & $478$ &  $1.18(1)$ & $2.0(2)$  & $0.93(4)$& $3$ &  ${\cal O}_{12345}$ & $9-13$ & $0.029$\\
\hline
$1/2$ & $3$ & $230$ &  $1.23(7)$ & $1.7(8)$  &          & $3$ &  ${\cal O}_{12345}$ & $7-10$ & $0.014$\\
$1/2$ & $3$ & $342$ &  $1.29(9)$ & $1.0(7)$  &          & $3$ &  ${\cal O}_{12345}$ & $8-10$ & $0.096$\\
$1/2$ & $2$ & $478$ &  $1.47(4)$ & $1.7(5)$  &          & $3$ &  ${\cal O}_{12345}$ & $8-10$ & $0.22$\\
\hline
\hline
$3/2$ & $1$ & $230$ &  $0.771(9)$& $2.1(2)$  & $2.2(1)$ & $3$ &  ${\cal O}_{123}$   & $9-13$ & $0.0062$\\
$3/2$ & $1$ & $342$ &  $0.937(6)$& $2.3(1)$  & $1.93(7)$& $3$ &  ${\cal O}_{123}$   & $9-13$ & $0.0047$\\
$3/2$ & $1$ & $478$ &  $1.156(5)$& $3.45(2)$ & $2.15(7)$& $3$ &  ${\cal O}_{123}$   & $9-13$ & $0.046$\\
\hline
$3/2$ & $2$ & $230$ &  $1.19(29)$& $0.09(22)$&          & $3$ &  ${\cal O}_{123}$   & $8-10$ & $0.00014$\\
$3/2$ & $2$ & $342$ &  $1.27(11)$& $0.09(12)$&          & $3$ &  ${\cal O}_{123}$   & $9-12$ & $0.23$\\
$3/2$ & $2$ & $478$ &  $1.55(7)$ & $0.47(33)$&          & $3$ &  ${\cal O}_{123}$   & $9-11$ & $0.12$\\
\hline
\hline
\end{tabular}
\end{center}
\caption{ \small  Analogous to Table \ref{tab_results_dyn}, but for the quenched simulation  at the larger volume $V=16^3\times 28$ and $a^{-1}\simeq 1~$GeV. \label{tab_results_q}}
\end{table}

\begin{table}[h]
\begin{center}
\begin{tabular}{c | c c c c c }  
$|I, I_3\rangle$ & $|1,1\rangle$ &$|1,-1\rangle$ & $|1,0\rangle$ &$|\tfrac{1}{2},\tfrac{1}{2}\rangle$ &$|\tfrac{1}{2},-\tfrac{1}{2}\rangle$ \\
\hline
$\bar q \Gamma q^\prime$ & $\bar d \Gamma u$ & $-\bar u \Gamma d$ & $\tfrac{1}{\sqrt{2}}[\bar d \Gamma d-\bar u \Gamma u]$ & $\bar s \Gamma u$ & $\bar s \Gamma d$\\
\hline
\end{tabular}
\end{center}
\caption{ \small $\bar q \Gamma q^\prime$ with definite $|I, I_3\rangle$, which are used to build two-particle interpolators via Clebsh-Gordan coefficients in Appendix B.}\label{tab_CG}
\end{table}

\end{document}